\documentclass[12pt,preprint]{aastex}
\usepackage{graphics}
\shorttitle {The SEGUE Stellar Parameter Pipeline. II. Validation
from Galactic Globular and Open Cluster Observations}

\shortauthors{Lee et al.}

\begin{document}

\title{The SEGUE Stellar Parameter Pipeline. II. Validation with Galactic Globular and Open Clusters}

\author{Young Sun Lee, Timothy C. Beers, Thirupathi Sivarani}
\affil{Department of Physics \& Astronomy, CSCE: Center for the
Study of Cosmic Evolution, and JINA: Joint Institute for Nuclear
Astrophysics, Michigan State University, East Lansing, MI 48824,
USA} \email{lee@pa.msu.edu, beers@pa.msu.edu, thirupathi@pa.msu.edu}

\author{Jennifer A. Johnson, Deokkeun An}
\affil{Department of Astronomy,\\ Ohio State University,
 Columbus, OH 43210}
\email{jaj@astronomy.ohio-state.edu,
deokkeun@astronomy.ohio-state.edu}

\author{Ronald Wilhelm}
\affil{Department of Physics,\\ Texas Tech University, Lubbock, TX
79409} \email{ron.wilhelm@ttu.edu}

\author{Carlos Allende Prieto, Lars Koesterke}
\affil{Department of Astronomy, \\ University of Texas, Austin, TX
78712} \email{callende@astro.as.utexas.edu}

\author{Paola Re Fiorentin}
\affil{Max Planck Institut f$\ddot{u}$r Astronomie, \\
K$\ddot{o}$nigstuhl 17, 69117 Heidelberg, Germany}
\email{fiorent@mpia-hd.mpg.de}

\author{Coryn A.L. Bailer-Jones}
\affil{Max Planck Institut f$\ddot{u}$r Astronomie, \\
K$\ddot{o}$nigstuhl 17, 69117 Heidelberg, Germany}
\email{calj@mpia-hd.mpg.de}

\author{John E. Norris}
\affil{Research School of Astronomy and Astrophysics, \\
Australian National University, Weston, ACT 2611, Australia}
\email{jen@mso.anu.edu.au}

\author{Brian Yanny}
\affil{Fermi National Accelerator Laboratory,\\
Batavia, IL 60510}
\email{yanny@fnal.gov}

\author{Constance Rockosi}
\affil{Department of Astronomy,\\
University of California, Santa Cruz, CA 95064}
\email{crockosi@ucolick.org}

\author{Heidi J. Newberg}
\affil{Department of Physics $\&$ Astronomy,\\
Rensselaer Polytechnical Institute, Troy, NY 12180}
\email{newbeh@rpi.edu}

\author{Kyle M. Cudworth}
\affil{Yerkes Observatory, \\
The University of Chicago, Williams
Bay, WI 53191} \email{kmc@yerkes.uchicago.edu}

\author{Kaike Pan}
\affil{Apache Point Observatory, \\
Apache Point Observatory, P. O. Box 59, Sunspot, NM 88349}
\email{kpan@apo.nmsu.edu}

\author{ }
\affil{}

\clearpage

\begin{abstract}

We validate the performance and accuracy of the current SEGUE (Sloan Extension
for Galactic Understanding and Exploration) Stellar Parameter
Pipeline (SSPP), which determines stellar atmospheric parameters (effective
temperature, surface gravity, and metallicity) by comparing derived overall
metallicities and radial velocities from selected likely members of three
globular clusters (M~13, M~15, and M~2) and two open clusters (NGC~2420 and
M~67) to the literature values. Spectroscopic and photometric data obtained
during the course of the original Sloan Digital Sky Survey (SDSS-I) and its
first extension (SDSS-II/SEGUE) are used to determine stellar radial velocities
and atmospheric parameter estimates for stars in these clusters. Based on the
scatter in the metallicities derived for the members of each cluster, we
quantify the typical uncertainty of the SSPP values, $\sigma (\rm [Fe/H])$ =
0.13 dex for stars in the range of 4500 K $\le T_{\rm eff} \le 7500$ K and $2.0
\le \log g \le 5.0$, at least over the metallicity interval spanned
by the clusters studied ($-2.3 \le {\rm [Fe/H]} < 0 $). The
surface gravities and effective temperatures derived by the SSPP are
also compared with those estimated from the comparison of the
color-magnitude diagrams with stellar evolution models; we find 
satisfactory agreement. At present, the SSPP underestimates
[Fe/H] for near-solar-metallicity stars, represented by members of
M~67 in this study, by $\sim$ 0.3 dex.

\end{abstract}

\keywords{methods: data analysis --- stars: abundances, fundamental parameters
          --- surveys ---  techniques: spectroscopic }

\section{Introduction}

The Sloan Extension for Galactic Understanding and Exploration (SEGUE) is one of
three key projects (LEGACY, SUPERNOVA SURVEY, and SEGUE) in the current
extension of the Sloan Digital Sky Survey, known collectively as SDSS-II. The
SEGUE program is in the process of obtaining $ugriz$ imaging of some 3500 square
degrees of sky outside of the SDSS-I footprint (Fukugita et al. 1996; Gunn et
al. 1998, 2006; York et al. 2000; Stoughton et al. 2002; Abazajian et al. 2003,
2004, 2005; Pier et al. 2003), with special attention being given to scans of
lower Galactic latitudes ($|b|$ $<$ 35$^{\circ}$) in order to better probe the
disk/halo interface of the Milky Way. SEGUE is also obtaining $R$ $\simeq$ 2000
spectroscopy over the wavelength range 3800 $-$ 9200\,{\AA} for some 250,000
stars in 200 selected areas over the sky available from Apache Point, New
Mexico.

The SEGUE Stellar Parameter Pipeline (hereafter, SSPP) processes the wavelength-
and flux-calibrated spectra generated by the standard SDSS spectroscopic
reduction pipeline (Stoughton et al. 2002), obtains equivalent widths and/or
line indices for 77 atomic or molecular absorption lines, and estimates $T_{\rm
eff}$, log $g$, and [Fe/H] through the application of a number of approaches.
The current techniques employed by the SSPP include a minimum distance method
(Allende Prieto et al. 2006), neural network analysis (Bailer-Jones 2000;
Willemsen et al. 2005; Re Fiorentin et al. 2007), auto-correlation analysis
(Beers et al. 1999), and a variety of line index calculations based on previous
calibrations with respect to known standard stars (Beers et al. 1999; Cenarro et
al. 2001a,b; Morrison et al. 2003). The SSPP employs five different methods for
estimation of $T_{\rm eff}$, eight for estimation of log $g$, and nine for
estimation of [Fe/H]. Details of the methods used are discussed in detail by Lee
et al. (2007a, hereafter Paper I). The use of multiple methods allows for
empirical determinations of the internal errors for each parameter, based on the
range of reported values -- typical internal errors for stars in the temperature
range $4500$~K $\le$ $T_{\rm eff}$ $\le$ 7500~K are $\sim$ 73~K, $\sim$ 0.19
dex, and $\sim$ 0.10 dex, in $T_{\rm eff}$, log $g$, and [Fe/H], respectively.
Allende Prieto et al. (2007, hereafter Paper III) point out that the internal
uncertainties provided by the SSPP underestimate the typical random errors at
high signal-to-noise ($S/N$) ratios because most methods in the SSPP make use of
similar parameter indicators (e.g., hydrogen lines for effective temperature)
and similar atmospheric models. Paper III empirically determines empirically external
uncertainties of $\sim$ 130~K, $\sim$ 0.21 dex, and $\sim$ 0.11 dex, for $T_{\rm
eff}$, log $g$, and [Fe/H], respectively, by comparison with high-resolution
spectroscopy ($7000 < R < 45,000$) of brighter SDSS-I/SEGUE that have been
obtained with 8m$-$10m class telescopes. Somewhat larger errors apply to stars
with temperatures near the extremes of the range above. The present study of
Galactic globular and open cluster stars tests the SSPP's ability to derive accurate
results for stars with a wide range of temperatures and gravities appropriate
for metal-poor and near-solar-metallicity stellar populations in the Galaxy, and
demonstrates that the derived metallicity scale is identical for dwarfs and
giants.

Although the SSPP will continue to evolve in the near future, it has been frozen
for now at the version used for obtaining results for stars with suitable data
from SDSS Data Release 6 (DR-6; Adelman-McCarthy et al. 2007b). Previous
versions of the SSPP have already been used for the analysis of SDSS-I
observations. For example, Allende Prieto et al. (2006) report on the
application of one of the methods included in the SSPP to some 20,000 F- and
G-type stars from SDSS-I DR-3 (Abazajian et al. 2005). Beers et al. (2006) have
compiled a list of over 6000 stars with [Fe/H] $< -2.0$ (including several
hundred with [Fe/H] $< -3.0$), based on application of the present SSPP to some
200,000 stars from SDSS-I DR-5 (Adelman-McCarthy et al. 2007a). Carollo et al.
(2007) reports on an analysis of the kinematics of relatively bright stars from
SDSS-I that have been used as calibration objects during the main survey.

In this paper, the second in the SSPP series, we show that estimates of the
atmospheric parameters and radial velocities obtained by the SSPP for
stars with a reasonable likelihood of membership in previously studied Galactic
globular and open clusters are sufficiently accurate to justify the use of the
present SSPP parameters for carrying out detailed studies of the halo and thick-disk
populations of the Milky Way. In deriving the overall iron abundance for each
cluster, we assume it comprises a chemically homogenous population.
%This assumption is valid since as Cohen \& Mel$\acute{\rm e}$ndez (2005) showed there
%has not been detected dispersion larger than the observational uncertainties in
%the heavy elements between Ca and the iron peak. Most of large chemical
%inhomogeneity occurs to the light elements such as C, N, O, Na, etc. and we are
%not comparing the abundance of these elements.

In \S 2, the photometric and spectroscopic data obtained for M~13, M~15, M~2,
NGC~2402, and M~67 are described. Section 3 presents the methods used to
separate likely cluster members from field stars in the directions toward these
clusters. Best estimates of the overall [Fe/H] and radial velocity of each
cluster are derived in \S 4. In \S 5 we compare the SSPP determinations of
$T_{\rm eff}$ and log $g$ for selected member stars in each cluster with their
expected positions on color-magnitude diagrams. A summary and brief conclusions
are provided in \S 6.

\section{Photometric and Spectroscopic Data}

Galactic globular and open clusters are nearly ideal testbeds for validation of
the stellar atmospheric parameters estimated by the SSPP. In most clusters, it
is expected that their member stars were born simultaneously out of well-mixed,
uniform-abundance gas at the same location in the Galaxy. Therefore, with the
exception of effects due to post main-sequence evolution, primordial variations
in carbon and nitrogen, or contamination from binary companions that have
transferred material, the member stars should exhibit very similar elemental
abundance patterns. Three of the clusters in our study, M~13, M~15, and M~2,
have well-known CN variations that extend to the main-sequence turnoffs (Smith
\& Briley 2006 for M13; Cohen, Briley, \& Stetson 2005 for M15; Smith \& Mateo
1990 for M2). However, these abundance variations can be ignored when deriving
metallicities from regions of the spectra that do not include CH, CN, or NH
features, as is the case with most of our techniques (those that may be affected
by the presence of such features are automatically de-selected in the
determination of the adopted [Fe/H]).

True cluster members should exhibit small radial velocity differences with
respect to their parent clusters. Furthermore, it is possible to examine
theoretical predictions of temperatures and surface gravities for member stars
that lie along the cluster main sequence (MS), red giant branch (RGB), or
horizontal branch (HB) in color-magnitude diagrams (CMDs). As part of tests of
the SEGUE star-selection algorithm (Adelman-McCarthy et al. 2007b) and the SSPP,
and during normal SEGUE operation, we have obtained $ugriz$ photometry and
medium-resolution (2.3 {\AA}; $R$ = 2000) spectroscopy for large numbers of
stars along lines of sight toward the globular clusters M~13, M~15, and M~2 and
the open clusters NGC~2420 and M~67. Below we discuss these photometric and
spectroscopic data in more detail.

\subsection{Photometric Data}

The SDSS obtains scans of the sky using the ARC 2.5m telescope on Apache Point,
New Mexico. These data are collected in five broad bands ($u, g, r, i, z$) with
central wavelengths 3551, 4686, 6166, 7480, and 8932\,{\AA} (Fukugita et al.
1996), respectively, using an imaging array of 30 ($6 \times 5$) 2048 $\times$
2048 Tektronix CCDs (Gunn et al. 1998). The pixel size is 24 $\mu$m,
corresponding to $0.396{''}$ on the sky. A series of software procedures,
collectively known as the SDSS PHOTO pipeline (Lupton et al. 2001), processes
and reduces the scanned images shortly after data are obtained. As part of these
procedures, the instrumental fluxes and astrometric positions (Pier et al.
2003), as well as a determination of whether an object is likely to be stellar
(i.e., a \emph{point source}), or not (an \emph{extended} source) are obtained.
Afterwards, the photometric data are further calibrated by matching to brighter
known standards observed with a smaller calibration telescope on Apache Point
(Hogg et al. 2001; Smith et al. 2002; Tucker et al. 2006). The processed
photometric data have been shown to exhibit 2\% relative and absolute errors
(0.02 magnitudes) in $g$, $r$, and $i$, and $3 \%-5 \%$ errors in $u$ and $z$
for all stellar objects brighter than $g = 20$ (Stoughton et al. 2002; Abazajian
et al. 2004, 2005; Ivezic et al. 2004). The first-pass photometric data for each
of the clusters used in the present study were secured by querying the DR-3
(Abazanjian et al. 2005), DR-5 (Adelman-McCarthy et al. 2007a), and DR-6
(Adelman-McCarthy et al. 2007b) releases from the SDSS Catalog Archive Server
(CAS).

Figure 1 illustrates one of the primary challenges in working with data for
clusters obtained with SDSS -- the automated PHOTO pipeline (Lupton et al. 2001)
was not designed to adequately deal with crowded fields such as the central
regions of globular clusters. As a result, essentially all of the stars in this
region (which are by definition the most likely ones to be cluster members) do
not have reported apparent magnitudes in the SDSS CAS. To circumvent this
limitation as much as possible, we have instead performed crowded field
photometry for the center of the clusters, using the DAOPHOT/ALLFRAME suite of
programs (Stetson 1987; Stetson 1994) in IRAF\footnote{IRAF is distributed by
the National Optical Astronomy Observatories, which is operated by the
Association of Universities for Research in Astronomy, Inc. under cooperative
agreement with the National Science Foundation.}. A full description of the
methods used and the photometric measures obtained is provided by Johnson et al.
(2007). Briefly, DAOPHOT was run on each image, and the five images of each
field (one for each filter) were then simultaneously run through ALLFRAME.
DAOGROW (Stetson 1990) was used to derive aperture corrections to the
point-spread-function photometry for the SDSS aperture radius of 7.4 arcsecs.
Finally, the zeropoint term from the {\it tsField} files was applied to calibrate the
data. This procedure also permits a check on the techniques used by the SDSS
PHOTO pipeline in regions outside the cluster where the areal density of sources
on the sky is sufficiently low that it may be used.

After completing the above procedures, we finally combine the results from the
PHOTO pipeline with those from the crowded-field photometry to obtain an almost
complete catalog of $ugriz$ photometry for stars in the region of each of our
program clusters. All photometric data are corrected for extinction and
reddening by application of the Schlegel, Finkbeiner, $\&$ Davis (1998) maps.
The average reddening ($E(B-V)$) for stars in the direction of these clusters is
0.017, 0.110, 0.045, 0.041, 0.032 for M~13, M~15, M~2, NGC~2420, and M~67,
respectively. Comparing with the literature values listed in Table 1, most of
the average reddenings of the clusters agree within about 0.02 mags.

\subsection{Spectroscopic Data}

The spectroscopy discussed in the present paper was obtained during the course
of SEGUE tests and normal SEGUE observations. In normal SEGUE operation mode, a pair of
plug-plates (referred to as the ``bright'' and ``faint'' plates) are obtained
over the $3^{\circ}$ field of the ARC 2.5m. A total of 640 optical fibers are
employed to obtain $R=2000$ spectra for on the order of 600 program
stars for each plate (the remaining fibers are used for spectrophotometric and
reddening calibration objects and observations of the night sky). The exposure
time depends on observation conditions. For a bright plate, exposures are set to
achieve a total $(S/N)^{2} >$ 15/1 from the two blue-side CCDs on the SDSS
spectrographs; the exposure for a faint plate is set such that a total $(S/N)
^{2} >$ 50/1 for all four (red and blue CCDs) on the SDSS spectrographs is
achieved. In order to identify and remove cosmic ray hits, each plate must have
at least three exposures; the integration time for any single exposure is not
longer than 30 minutes. For the purposes of targeting objects on these plates,
the boundary between the bright and faint plates is set at $r \sim$ 18.0. The
data thus obtained are processed through the SDSS spectroscopic pipeline
software (SPECTRO2D and SPECTRO1D), which produces wavelength and
flux-calibrated spectra, and also obtains estimates of radial velocities and
line indices (Stoughton et al. 2002). Tests of the quality of stellar radial
velocities from the SSPP (which uses initial estimates from the SDSS processing
pipelines) indicate precisions better than 5 km~s$^{-1}$ are achieved for
brighter stars, with zero-point offsets of no more than a few km~s$^{-1}$,
respectively (Paper III). These errors degrade for fainter stars, as expected.

An initial set of candidate member stars of the globular and open clusters
studied in the present paper were selected on the basis of photometric and
astrometric data (proper motions) from the literature. The central cores of the
clusters were not targeted because the PHOTO pipeline does not resolve the very
crowded fields into single star detections, and also due to limitations on the
separations of the fibers during the spectroscopic follow-up stage. The primary
method for selecting member candidates was performed by plotting a photometric
CMD for a given cluster, and choosing stars from regions of this diagram that
correspond to location on the MS turnoff or RGB of the cluster. An additional
list of bright stars for M~15 and M~2 with previously available proper motions
consistent with membership in the clusters was provided by Cudworth (1976 and
private communication) and Cudworth $\&$ Rauscher (1987). Other stars in the
fields of these clusters were used to fill spectroscopic fibers using the
default SEGUE target selection algorithm (Adelman-McCarthy et al. 2007b). While
many of these additional targets turned out to be stars from the general field
populations, a significant fraction turned out serendipitously to be members of
the clusters.

For M~13, three specially designed plates were obtained. Two of the three plates
followed the standard SEGUE target selection procedure (Adelman-McCarthy et al.
2007b) of sampling stars with a variety of spectral types based on the SDSS
imaging and PHOTO processing. An additional set of likely M~13 members,
including several stars that were saturated in the SDSS image ($r < 14.5$) and
with coordinates from Cudworth $\&$ Monet (1979) and Cudworth (private
communication), were added to the target list with high priority (bumping
ordinary SEGUE targets), in order to obtain spectra of several likely giant-branch and
horizontal-branch members.

In the case of NGC~2420, the stars chosen for spectroscopy were primarily
targeted from the SDSS photometry obtained by the PHOTO pipeline, using the
normal SEGUE target selection algorithm. Additional stars with apparent
magnitudes in the range 14.5 $< g <$ 20.5 that fell within 0.5 degrees
from the center of NGC~2420 were also targeted for spectroscopy. However, due to
crowding, if two objects were within 55${''}$ of one another, then only one
received a fiber. Thus, not every star in the central region of NGC~2420 was
targeted. There were about 480 objects selected in this way, including a number
of non-cluster members that are located in the NGC~2420 field.

For M~67, the initial targets came from the SDSS imaging data processed by the
PHOTO pipeline. However, for this cluster, many candidate members with
positions, magnitudes, and colors from the WEBDA (http:
//www.univie.ac.at/webda/) catalogs were added to the target lists. The bright
targets (with $r < 14$) saturate the SDSS imaging camera, so these were added
from the literature (Sanders 1989; Fan et al. 1996). Such bright stars normally
saturate a regular SDSS spectroscopic exposure, so there were exposed for
shorter than normal. About 200 very bright stars between about 12 $<$ $g$ $<$ 14
were targeted.

In total, we obtained SDSS spectroscopy for 1920, 1280, 640, 1280, and 640
targets, including sky spectra and calibration object spectra, in the fields of
M~13, M~15, M~2, NGC~2420, and M~67 respectively. The reduced spectra were then
processed through the SSPP in order to estimate $T_{\rm eff}$, log $g$, and
[Fe/H], among other quantities. Table 1 summarizes the global properties of the
clusters under consideration in this paper, taken from the compilation of Harris
(1996) for the globular clusters and from WEDBA or Gratton (2000) for the open
clusters.

\subsection{Radial Velocities}

There are two estimated radial velocities provided from the SDSS spectroscopic
pipeline. One is an absorption redshift obtained by cross-correlating the
spectra with templates that were obtained from SDSS commissioning spectra
(Stoughton et al. 2002). Another comes from matching the spectra with ELODIE
template spectra (Prugniel $\&$ Soubiran 2001). In most cases the velocity based
on the ELODIE template matches appears to be the best available estimate, as
spectra of ``quality assurance'' stars with multiple measurements show the most
repeatable values for this estimator. However, this is not always the case. We
proceed to select the best available velocity in the following manner. If the
velocity determined by comparison with the ELODIE templates has a reported error
of 20 km s$^{-1}$ or less then this velocity is adopted. If the error from the
ELODIE template comparison is larger than 20 km s$^{-1}$, and the relative
offset between the two radial velocities is less than 40 km s$^{-1}$, we take an
average of the two. If none of the criteria above are satisfied, which happens
only rarely, and mainly for quite low $S/N$ spectra, or for hot/cool stars
without adequate templates, we obtain the calculated radial velocity from a
custom routine that examines the wavelengths of a number of prominent absorption
features. If none of these methods yield a reasonable estimate of radial
velocity, or it appears spurious (i.e., falls outside of the range $\pm 1000$ km
s$^{-1}$), we simply ignore the star in subsequent analyses. A more detailed
description of the procedures used for determination of the best available
radial velocity, and of zero-point offsets of the radial velocities, can be found
in Paper I.

\section{Membership Selection from the Spectroscopic Samples}

Owing to an insufficient number of stars with available spectroscopy for each
cluster, it is not possible to obtain a well-defined Color Magnitude Diagram
(CMD) based solely on spectroscopically confirmed member stars. Thus, we make
use of photometric data in the field of each cluster, and describe below how we
obtain a relatively clean CMD for individual clusters, and select likely member
stars from the spectroscopic data.

\subsection{Likely Member Star Selection for Globular Clusters}

One of the primary issues that one needs to address when creating a CMD, or
selecting likely member stars, for a star cluster is removal of contamination
from field stars. In order to approximately isolate the likely cluster members
from the field stars we have made use of the CMD mask algorithm described by
Grillmair et al. (1995). We illustrate the basic idea by application of this
algorithm to the M~13 field shown in Figure 1. We first select all stars inside
the estimated tidal radius (25.2$^{'}$; Harris 1996), shown as the innermost
green circle in Figure 1. This is regarded as the cluster region. The red dots
represent stars with available photometry from the SDSS PHOTO pipeline (Lupton
et al. 2001); the black dots are stars with photometry obtained from DAOPHOT.
The blue open circles indicate stars with available spectroscopy. We then choose
an annulus outside the cluster region, indicated on the Figure as the region
between the two black circles, as the field or background region.

We next obtain CMDs of each region, spanning $-1.0 \le (g-r)_0 \le 1.5$ and $12
\le g_0 \le 22$, and then subdivide these diagrams such that the size of each
sub-grid is 0.2 mag wide in $g_0$ and 0.05 mag wide in $(g-r)_0$ color. The
total number of sub-grids for the CMDs in each region is thus 2500
(50$\times$50). Figure 2 shows the resulting CMDs of the cluster (left panel)
and field (right panel) regions, overplotted with squares representing the
selected sub-grids, obtained as described below.

We first calculate the signal-to-noise ($s/n$) in each preliminary sub-grid by
application of Eqn. (1) over the entire CMD region shown in Figure 2. Here we
assume that the field stars outside the tidal radius are uniformly distributed
throughout the annulus area.

\begin{equation} s/n(i,j) = \frac{n_{c}(i,j) - gn_{f}(i,j)}{\sqrt{n_{c}(i,j) + g^{2}n_{f}(i,j)}}.
\label{eq1}
\end{equation}

\noindent In the above, $n_{c}$ and $n_{f}$ refer to the number of stars in
each sub-grid with color index $i$ and magnitude index $j$, counted within the
cluster region and field region, respectively. The parameter $g$ represents the
ratio of the cluster area to the field area.

The following procedures are followed in order to find the optimal range of
colors and magnitudes that correspond to the likely members of each cluster.
First, we sort the elements of $s/n(i,j)$ in descending order, so that we obtain
a one-dimensional array of $s/n(i,j)$ with index $l$; the array element with the
highest $s/n(i,j)$ corresponds to $l=1$. The next step is to obtain star counts
in gradually larger regions of the CMDs. The accumulated area is represented as
$a_{k} = ka_{l}$, where $a_{l} = 0.01$ mag$^{2}$, which is the same for all
sub-grids, and is the area of a single sub-grid in the CMD array, and the $k$ is
the number of sub-grids to combine. Finally, the cumulative signal-to-noise
ratio, $S/N(a_{k}) $, as a function of $a_{k}$, is calculated from:

\medskip
\begin{equation} S/N(a_{k}) = \frac{N_{c}(a_{k}) - gN_{f}
(a_{k})}{\sqrt{N_{c} (a_{k})  + g^{2} N_{f} (a_{k})}}
\label{eq3}
\end{equation}
\medskip

\noindent where,

\begin{equation}
 N_{c}(a_{k}) = \sum_{l=1}^{k} n_{c}(l), ~~~N_{f}(a_{k}) = \sum_{l=1}^{k} n_{f}(l) \label{eq4}
\end{equation}
\medskip

\noindent The parameter $n_{c}(l)$ denotes the number of stars within the
cluster region having ordered color-magnitude index $l$; $n_{f}(l)$ represents
the same quantity for the field region. Based on the maximum value of
$S/N(a_{k})$, a threshold value of $s/n$ is picked in order to select high-
contrast surface-density areas (i.e, high $s/n$) between the cluster and field
regions. These are considered to be the sub-grids that contain likely cluster
members. After removing single-star events in areas of the CMDs where the
field-star density is low, all stars in sub-grids with $s/n(i,j)$ greater than
the threshold value of $s/n$ are selected. These stars are considered as the
photometrically likely member stars for a given cluster.

The red squares shown in the left panel of Figure 2 are the sub-grids with $s/n$
greater than the threshold value; the corresponding sub-grids in the field region
are shown as green squares in the right panel of this Figure. Figure 3 depicts
the CMD of the selected likely members of M~13 from the photometric data, shown
as black dots. The same procedures are performed to differentiate the likely
member stars of M~15 and M~2 from the photometric sample.

We now proceed to select the stars that are likely members of the globular
clusters from the available spectroscopic sample. This step begins by selection
of the stars within the cluster tidal radii that pass the photometric criterion
for membership, based on their location on the cluster CMDs according to the
algorithm described above.

Figure 3 displays the cleaned CMD of M~13, overplotted with the likely members
from the spectroscopic sample (shown as red circles). The same procedures are
carried out to identify likely member stars of M~15 and M~2 from their
spectroscopic data. Based on these cuts, at this stage of the analysis there are
296 (338) likely members for M~13, 124 (160) for M~15, and 21 (22) for M~2
identified. In the above, the first listed numbers indicate the stars with
available estimates of [Fe/H] from the SSPP, while the quantities in parentheses
represent the number of stars with available radial velocities (RVs). Additional
cuts, based on the derived metallicity estimates and RVs, are described in \S 4.

\subsection{Likely Member Star Selection for Open Clusters}

Since the fields of nearby open clusters are not as crowded as those of globular
clusters, the signal-to-noise ratio between the cluster region and the
background region is not sufficiently high to select likely cluster members by
means of the CMD mask algorithm. As an alternative, we first obtain a fiducial
line for an open cluster (including its main sequence and sub-giant branch, if
it exists) by use of an robust polynomial fitting procedure. As an example,
Figure 4 shows the CMD of the NGC~2420 field inside a radius of 0.3 degrees from
the center of the cluster. According to the open cluster catalog of Dias et al.
(2002), the apparent diameter of this cluster is only 5$^{'}$ on the sky, but we
prefer to adopt a 20$^{'}$ radius, in order to include as many member stars as
possible. The red line is the fiducial line derived from the robust polynomial
fit. The blue lines are the upper and lower limits (fiducial $\pm$ 0.06 dex in
$(g-r)_{0}$), determined by eye. Stars from the spectroscopic sample that fall
within the 20$^{'}$ radius and inside the blue limit lines in Figure 4 are
identified.

A similar procedure is applied to M~67, except that a 30$^{'}$ (apparent
diameter of 25$^{'}$; Dias et al. 2002) radius and fiducial $\pm$ 0.10 mag in
$(g-r)_{0}$ is used. Based on this selection method, there are 195 (234) and 61
(64) for NGC~2420 and M~67, respectively. The first listed numbers indicate the
stars with available estimates of [Fe/H] from the SSPP, while the quantities in
parentheses represent the number of stars with available radial velocities (RVs)
. Additional cuts, based on the derived metallicity estimates and RVs, are
described below.

\section{Determination of Overall Metallicities and Radial Velocities of the
Clusters}

In order to investigate the accuracy of our derived metallicities and RVs, we
now consider the global distribution of these parameters obtained from the
current version of the SSPP for the likely cluster members. In this section, we
describe a method to best isolate ``true member stars'' from the spectroscopic
samples described above. We then use these subsamples to determine our best
estimates of the overall metallicities and RVs of the clusters considered in
this study.

\subsection{Selection of True Members}

We establish the criteria for carrying out metallicity and RV cuts as follows.
%Note that as mentioned in \S 1, because there is no manifest evidence of large
%dispersion in the abundance of the iron peak elements (e.g., Cohen \&
%Mel$\acute{\rm e}$ndez 2005) for the clusters currently under consideration. we
%suppose the clusters have chemically mono population and make use of the
%metallicity as a membership criterion.

The left panel of Figure 5 illustrates the [Fe/H] distribution for three
different subsamples of stars. The first, shown as the black dot-dashed line,
represents the distribution of derived metallicities for the 1547 stars with
available estimates of [Fe/H] along the line of sight to M~13. Note that this
distribution includes numerous stars that cannot be considered members of the
cluster, as they cover a much wider range of [Fe/H] than might be expected if
they were drawn exclusively from the cluster member population. The dot-dashed
line in the right panel of Figure 5 shows the RV distribution of these same
stars.

The red dashed line in the left panel of Figure 5 is the
distribution of [Fe/H] for the 296 likely members selected from the
spectroscopic sample as described above. We obtain a Gaussian fit to
the {\it highest peak} of the distribution of these stars (solid
blue line in Figure 5), and obtain an estimate of the mean
($<$[Fe/H]$>$) and standard deviation ($\sigma$) for this
distribution. Similar fits are obtained for the distribution of RVs
for the likely members shown in the right panel of Figure 5. On the
basis of these fits, we now trim likely outliers by application of a
2-$\sigma$ clipping procedure, for example:

\medskip
\begin{equation}
\rm <[Fe/H]> - 2\sigma_{\rm [Fe/H]} \leq \rm [Fe/H]_{\star} \leq
\rm<[Fe/H]> + 2\sigma_{\rm [Fe/H]} \label{eq6}
\end{equation}

\medskip
\begin{equation}
\rm <RV> - 2\sigma_{\rm RV} \leq \rm RV_{\star} \leq \rm<RV> +
2\sigma_{\rm RV} \label{eq7}
\end{equation}

\noindent In the above, [Fe/H]$_{\star}$ and RV$_{\star}$ correspond to the
values of these parameters for each star under consideration. The stars
surviving both of these clips are considered true cluster members for the
purpose of this study. Note that at no point have we considered the external
``known'' values of [Fe/H] and RV for the clusters as a whole.

Based on the application of these membership cuts, we now have a total of 169
stars identified as true members of M~13, 63 stars as true members of M~15, 9
stars as true members of M~2, 195 stars as true members of NGC~2420, and 51
stars as true members of M~67. The distribution of [Fe/H] and RV for the
surviving members of M~13 are shown as green histograms in the left and right
panels of Figure 5, respectively. Similar plots for M~15, M~2, NGC~2420, and
M~67 are shown in Figures 6, 7, 8, and 9, respectively. The distribution of
[Fe/H] for the selected true member stars of each cluster, as a function of
$T_{\rm eff}$, is shown in Figure 10.

Table 2 summarizes the results of the above exercise. Column (1) lists the
cluster name. Columns (2) and (3) are the lower and upper limits for the
2-$\sigma$ cuts on [Fe/H], respectively. Columns (4) and (5) are the
corresponding adopted limits on RV used for these cuts.

Tables 4$-$8 list the observed and derived quantities for all of the
individual stars considered as true member stars in the analysis of
each cluster. The columns are as defined in the table notes for
Table 4.

\subsection{Determination of Overall Estimates of Mean
Cluster Metallicity and Radial Velocity}

We now obtain final estimates of the cluster metallicities and RVs based on
Gaussian fits to the surviving true member stars for each cluster, as shown by
the blue curves in Figures 5$-$9. Table 2 summarizes these determinations. The
mean metallicity and 1-$\sigma$ spread of the metallicities of the true member
stars are listed in columns (6) and (7), respectively. Similar quantities for
the RVs are listed in columns (8) and (9). Column (10) lists the total number of
true member stars associated with each cluster, based on our analysis. External
estimates of the metallicities and RVs for these clusters, adopted from the
Harris (1996) compilation for M~13, M~15, and M~2 and from WEBDA (and references
therein) for NGC~2420 and M~67, are listed in columns (11) and (12). Column (13)
lists metallicity estimates for these clusters obtained from high-resolution
spectroscopy of a limited number of brighter stars by Kraft \& Ivans (2003) for
M~15 and M~13, Ivans (private communication) for M~2, and Gratton (2000) for
NGC~2420 and M~67.

For M~13, our estimate of the mean abundance, $<$[Fe/H]$>$ = $-1.56$, is very
close to the Harris (1996) estimate ([Fe/H]$_{\rm H}$ = $-$1.54). However, the
recent study of Kraft \& Ivans (2003) reported a revised cluster abundance for
M~13, derived from high-resolution spectroscopy of 28 giants. Their value
indicates a metallicity for M~13 that is a bit lower than that given by Harris,
[Fe/H]$_{\rm HR}$ = $-$1.63, and is lower by 0.07 dex than our estimate. Cohen
\& Mel$\acute{\rm e}$ndez (2005) reported [Fe/H] = $-1.50$ from a
high-resolution ($R$ = 35,000) analysis of a sample of 25 stars, consisting of
stars from the giant branch to near the main-sequence turnoff. Our derived
spread in the metallicities of the M~13 true member stars (0.16 dex) is also
satisfyingly low, especially considering the wide range of temperatures for true
members that are considered here. Our estimate of the mean radial velocity,
$<$RV$>$ = $-$245.1 km s$^{-1}$, with a standard deviation of 8.7 km s$^{-1}$,
is in good agreement with that given by Harris ($-$245.6 km s$^{-1}$). It is
important to note that, as mentioned in Paper I, we have already
added $+$7.3 km s$^{-1}$ to all DR-6 (Adelman-McCarthy et al. 2007b) stellar
radial velocities. Before the adjustment of this offset, an average offset of
$-$8.6 km s$^{-1}$ for M 13 and $-$6.8 km s$^{-1}$ for M 15 is obtained. Thus,
together with an offset of $-$6.6 km s$^{-1}$ that was obtained from a
preliminary result of a high-resolution spectroscopic analysis of SDSS-I/SEGUE
stars (Paper III) before DR-6 (Adelman-McCarthy et al. 2007b), we derived an
average offset of $-$7.3 km s$^{-1}$. However, a recent high-resolution
spectroscopic analysis of SDSS-I/SEGUE stars indicates an offset of about $-$6.9
km s$^{-1}$, resulting in an average of $-$7.4 km s$^{-1}$, Hence, in future data
releases (e.g, DR-7), this very minor difference might be reflected. For the
analysis of our clusters, all radial velocities have been corrected by $+$7.3 km
s$^{-1}$, in order to be consistent with DR-6.

Our estimate of the mean abundance of M~15, $<$[Fe/H]$>$ = $-2.12$, is close to
the value listed by Harris ([Fe/H]$_{\rm H}$ = $-$2.26). While Kraft $\&$ Ivans
(2003) obtained [Fe/H]$_{\rm HR}$ = $-$2.42, based on high-resolution
spectroscopy for nine giants in this cluster, Otsuki et al. (2006) reported
[Fe/H] = $-$2.29 from an analysis of high-resolution spectra for six giants
belonging to this cluster. Our derived spread in the metallicities of true
member stars in M~15 is quite low (0.14 dex). Our estimate of the mean radial
velocity, $<$RV$>$ = $-$107.4 km s$^{-1}$, with a standard deviation of 10.5 km
s$^{-1}$, agrees very well with that of the Harris (1996) value ($-$107.0 km
s$^{-1}$).

There are only a very small number of true member stars (9) for M~2. Their
average metallicity, $<$[Fe/H]$>$ = $-$1.58, is similar to the Harris (1996)
value ([Fe/H]$_{\rm H}$ = $-$1.62), and is in very good agreement with the value
obtained by Ivans (private communication) ([Fe/H]$_{\rm HR}$ = $-$1.56). The
estimated spread in our derived metallicities, 0.08 dex, is quite small. Our
estimate of the mean radial velocity, $<$RV$>$ = $+$0.4 km s$^{-1}$, with a
standard deviation of 7.7 km s$^{-1}$, is higher (by about 6 km s$^{-1}$) than
that provided by Harris ($-$5.3 km s$^{-1}$). Clearly, for the purposes of
validation of the SSPP, it would be highly desirable to obtain a larger number
of member stars in M~2; a new plate (640 spectra) will be obtained in the near
future. Note that M~2 presents a special challenge, since its mean metallicity
is quite close to that expected for members of the field halo population, while
its radial velocity is buried in the peak of foreground disk stars. Our
stringent criterion for true cluster members should remain effective, however,
since few non-cluster members will fulfill both the RV and metallicity criteria.

There are 130 true member stars selected for the open cluster NGC~2420. The mean
iron abundance of the selected true member stars is $<$[Fe/H]$>$ = $-$0.46,
which is in excellent agreement with the value ([Fe/H] = $-$0.44) determined by
Gratton (2000) from high-resolution spectroscopy of one member star. Friel $\&$
Janes (1993) reported [Fe/H] = $-$0.42 for nine member stars, based on medium-
and low-resolution spectroscopic data. Friel et al. (2002) determined [Fe/H] =
$-$0.38 $\pm$ 0.07, based on medium-resolution spectra of 20 member stars. Most
of these literature values are within the spread of our derived value. The
derived spread in the metallicities of true member stars (0.12 dex) is very low.
The radial velocity for NGC~2420 listed in Table 2, ($+74.0$ km s$^{-1}$), is an
average of the values $+84.0$ km s$^{-1}$, $+71.1$ km s$^{-1}$, and $+67.0$ km
s$^{-1}$ from Friel (1989), Scott et al. (1995), and Rastorguev et al. (1999),
respectively. This value agrees very well with our derived estimate of $+74.8$
km s$^{-1}$, with a standard deviation of $6.2$ km s$^{-1}$.

For M 67, 51 stars are identified as true member stars. A mean metallicity of
$<$[Fe/H]$> = -$0.35 is derived, with a small spread of 0.15 dex. This derived
$<$[Fe/H]$>$ differs by 0.37 dex from that of Gratton (2000), [Fe/H] = $+$0.02,
who derived this value from a high-resolution study of one member star. Randich
et al. (2007) analyzed 10 member stars of this cluster, based on high-resolution
($R \sim 45,000$) spectroscopy, and derived [Fe/H] = $+0.03 \pm 0.01$. [Fe/H] =
$+0.02 \pm 0.03$ was determined by Yong et al. (2005) from a high-resolution
spectroscopic analysis of three member stars. However, based on
medium-resolution spectra of 25 members, Friel et al. (2002) reported [Fe/H] =
$-0.15 \pm 0.05$. Other catalogs of open clusters (e.g., Twarog et al. 1997;
Chen et al. 2003) also report a solar metallicity for this cluster. The
literature values based on high-resolution spectroscopic analyses clearly
suggest that the present SSPP tends to under-estimate [Fe/H] by about 0.3 dex
for near-solar-metallicity stars. This is perhaps related to the difficulties
arising from the strong atomic and molecular lines (and possibly unreliable
synthetic spectra) for metal-rich stars. As mentioned by Gratton (2000), it
might be desirable to re-calibrate the metallicity scale used for the analysis
of medium-resolution spectra for stars with the solar iron abundance to better
match the results obtained from high-resolution analyses. The radial velocity of
$+32.9$ km s$^{-1}$ for M 67 from the literature listed in Table 2 (the average
of the Scott et al. 1995; Friel $\&$ Janes 1993; Rastorguev et al. 1999 values)
agrees with our derived mean radial velocity of $+34.9$ km s$^{-1}$ within the
standard deviation of our measurements, 5.6 km s$^{-1}$.

Thus, taking into account only the scatter in the metallicities and radial
velocities calculated from the members of each cluster, we are able to derive
estimates of typical external uncertainties for the SSPP values, $\sigma (\rm [Fe/H])$ =
0.13 dex, and $\sigma (\rm RV)$ = 7.7 km~s$^{-1}$ (after 5$\sigma$ clipping is
applied).

\section{A Comparison of Derived $T_{\rm eff}$ and log $g$ for
True Cluster Members with Color-Magnitude Diagrams}

In the previous section, we have considered the accuracy with which the SSPP
obtains estimates of metallicity and radial velocity. We now consider the
accuracy with which the SSPP obtains estimates of effective temperatures,
$T_{\rm eff}$, and surface gravities, log $g$. One excellent ``global'' test of
these estimates is to examine the locations of the true member stars on the
observed CMD (based on the totality of likely photometric member data) for each
cluster. One can also compare with corresponding theoretical CMDs.

Figures 11$-$15 show plots of the SSPP-estimated temperatures and gravities for
true member stars superposed on the photometrically cleaned CMDs for each of our
clusters. Note that in order to obtain the theoretical temperature scales (shown
along the top of the left-hand panels in each Figure), we make use of a linear
relation between $(g-r)_0$ color and $T_{\rm eff}$ by performing a least square
fit in this plane to the theoretical models of Girardi et al. (2004). We choose
the isochrones from this study that have the closest [Fe/H] to the derived
metallicity of each cluster, and adopt an age of 13.5 Gyr for the globular
clusters, 2.2 Gyr for NGC 2420, and 4.3 Gyr for M 67 (adopting the ages from
WEBDA for the open clusters). A similar procedure is applied for transforming the
$g_0$ magnitude to a theoretical log $g$ scale (shown along the far right axis
in the right-hand panels of each Figure). Distance moduli from the Harris (1996)
compilation for the globular clusters and WEBDA for the open clusters (also
listed in Table 1 of this study) are used in order to compute apparent
magnitudes.

In these Figures, we plot the SSPP-estimated parameters for true member stars in
different colors, corresponding to different ranges of temperature and surface
gravity (as shown in the legend for each plot). Each color represents a range of
500~K in $T_{\rm eff}$ and 0.5 dex in log $g$. The effective temperature
estimated by the SSPP appears in excellent agreement for most of the true member
stars, with only a few exceptions. Such stars could either be outright errors in
SSPP predictions, or could just be foreground/background stars that survived the
various membership cuts we have applied. Inspection of the Figures also reveals
the presence of a few stars close to the main-sequence TO in M~13 and M~15 that
appear to have slightly lower SSPP-estimated log $g$ than expected from the
theoretical scale. The surface gravities of stars along the RGB appear to be
very well estimated. Such behavior is perhaps to be expected, since the stars
close to the TO region are at the low end of the $S/N$ range that we accept for
analysis, and thus are subject to greater errors in the determination of their
atmospheric parameters. The RGB stars are among the brightest, and hence are
likely to have the best-determined estimates.

Inspection of Figure 14 for NGC 2420 indicates that gravity estimates for
most of the main-sequence stars are well-estimated from the SSPP, with the
exception of the faintest stars. These stars have only low $S/N$ spectra
available, resulting in higher uncertainties in determinations of their surface
gravities. It should also be recalled that surface gravity is a difficult
parameter to estimate, especially from spectra of the resolving power obtained
by the SDSS. Overall, we are pleased to see as good a behavior in the estimates
of this parameter as is demonstrated in Figures 11$-$15.

In addition, using the derived relations between $(g-r)_0$ and $T_{\rm eff}$,
and $g_0$ and log~$g$ from the isochrones, we predicted $T_{\rm eff}$ and
log~$g$ from the observed $(g-r)_0$ and $g_0$, respectively. Table 3 lists the
averages and standard deviations of the residuals of the effective temperatures
and surface gravities between the SSPP estimates and the calculated values. Even
though we have employed a simple relationship between $(g-r)_0$ with $T_{\rm
eff}$, we see good agreement between the SSPP estimates and the theoretical
values in $T_{\rm eff}$. Although, as expected, we notice a rather large offset
and scatter in the gravity, indicating a more complex function is needed, the
scatters are still within each bin size (0.5 dex) in Figures 11$-$15. M~2
exhibits a larger scatter in both $T_{\rm eff}$ and log~$g$ than the other
clusters, owing to the small number of member stars selected.

\section{Summary and Conclusions}

Based on photometric and spectroscopic data reported in SDSS-I and
SDSS-II/SEGUE, we have examined estimates of stellar atmospheric parameters and
heliocentric radial velocities obtained by the SEGUE Stellar Parameter
Pipeline (SSPP) for likely members of three Galactic globular clusters, M~13,
M~15, and M~2, and two open clusters, NGC~2420 and M~67, and compared them with
those obtained by external estimates for each cluster as a whole.

From the derived scatters in the metallicities and radial velocities obtained
for the likely members of each cluster, we quantify the typical external
uncertainties of the SSPP-determined values, $\sigma (\rm [Fe/H])$ = 0.13 dex,
and $\sigma (\rm RV)$ = 7.7 km~s$^{-1}$, respectively. These uncertainties apply
for stars in the range of 4500 K $\le T_{\rm eff} \le 7500$ K and $2.0 \le \log
g \le 5.0$, at least over the metallicity interval spanned by the clusters
studied ($-2.3 \le {\rm [Fe/H]} < -0.4$). Therefore, the metallicities and
radial velocities obtained by the SSPP appear sufficiently accurate to be used
for studies of the kinematics and chemistry of the metal-poor and moderately
metal-rich stellar populations in the Galaxy. We have also confirmed that
$T_{\rm eff}$ and log $g$ are sufficiently well-determined by the SSPP to
distinguish between different luminosity classes through a comparison with
theoretical predictions.

A comparison of the analysis of the available high-resolution spectroscopy of
SDSS-I/SEGUE stars (Paper III) with the SSPP predictions indicates that the
uncertainty in radial velocities adopted by the SSPP is no more than 5
km~s$^{-1}$ (after adjusting for an empirical offset of +7.3 km~s$^{-1}$). The
empirically determined precisions in estimated atmospheric parameters are $\sim$
130~K for effective temperature, $\sim$ 0.21 dex for surface gravity, and $\sim$
0.11 dex for [Fe/H]. These errors apply to the brightest stars obtained by
SDSS-I/SEGUE observations, on the order of $14.0 \le g \le 15.5$, and are
expected to degrade somewhat for fainter stars. We also found that the SSPP
tends to underestimate [Fe/H] for near-solar-metallicity stars (represented by
members of M~67 in this study), by $\sim$ 0.3 dex.

In future papers we will compare the predictions of the SSPP with
intermediate-metallicity clusters ([Fe/H] $\sim$ $-$0.7) and with additional
near-solar-metallicity populations, as sampled by metal-rich globular clusters
and nearby open clusters. Additionall metal-poor clusters will also be examined.
Further refinements in the SSPP, which hopefully will be better able to
recover accurate abundances for near-solar-metallicity stars, are anticipated.

\acknowledgements

Funding for the SDSS and SDSS-II has been provided by the Alfred P.
Sloan Foundation, the Participating Institutions, the National
Science Foundation, the U.S. Department of Energy, the National
Aeronautics and Space Administration, the Japanese Monbukagakusho,
the Max Planck Society, and the Higher Education Funding Council for
England. The SDSS Web Site is http://www.sdss.org/.

The SDSS is managed by the Astrophysical Research Consortium for the
Participating Institutions. The Participating Institutions are the
American Museum of Natural History, Astrophysical Institute Potsdam,
University of Basel, University of Cambridge, Case Western Reserve
University, University of Chicago, Drexel University, Fermilab, the
Institute for Advanced Study, the Japan Participation Group, Johns
Hopkins University, the Joint Institute for Nuclear Astrophysics,
the Kavli Institute for Particle Astrophysics and Cosmology, the
Korean Scientist Group, the Chinese Academy of Sciences (LAMOST),
Los Alamos National Laboratory, the Max-Planck-Institute for
Astronomy (MPIA), the Max-Planck-Institute for Astrophysics (MPA),
New Mexico State University, Ohio State University, University of
Pittsburgh, University of Portsmouth, Princeton University, the
United States Naval Observatory, and the University of Washington.

Y.S.L., T.C.B., and T.S. acknowledge partial funding of this work
from grant PHY 02-16783: Physics Frontiers Center / Joint Institute
for Nuclear Astrophysics (JINA), awarded by the U.S. National
Science Foundation. NASA grants (NAG5-13057, NAG5-13147) to C.A.P.
are thankfully acknowledged. J.E.N acknowledges support from
Australian Research Council Grant DP0663562. C.B.J and P.R.F
acknowledge support from the Deutsche Forschungsgemeinschaft (DFG)
grant BA2163.

\clearpage

%% Figure 1
\begin{figure}
\centering
\plotone{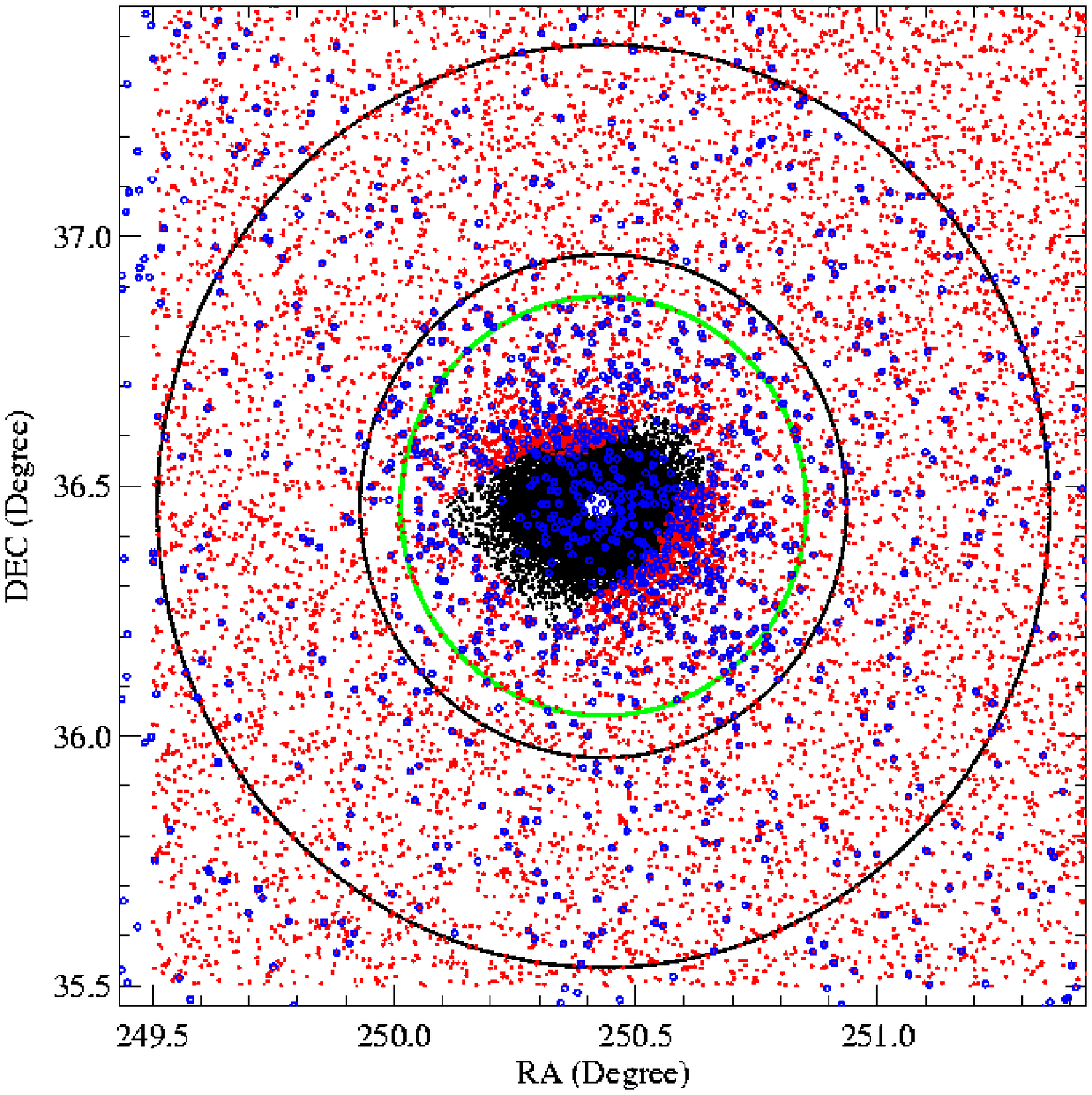}
%\plotone{f1.eps}
\caption{Stars with
available photometry in the field of M~13. The red dots represent
photometry from the SDSS PHOTO pipeline, while the black dots are
from the crowded-field photometry analysis. Blue open circles
indicate stars with available SDSS spectroscopy. The green circle is
the tidal radius. The region inside this radius is regarded as the
cluster region; the annulus between the two black circles is
considered the field region.}
\end{figure}
\clearpage

%% Figure 2
\begin{figure}
\centering
\includegraphics[angle=90,scale=0.60]{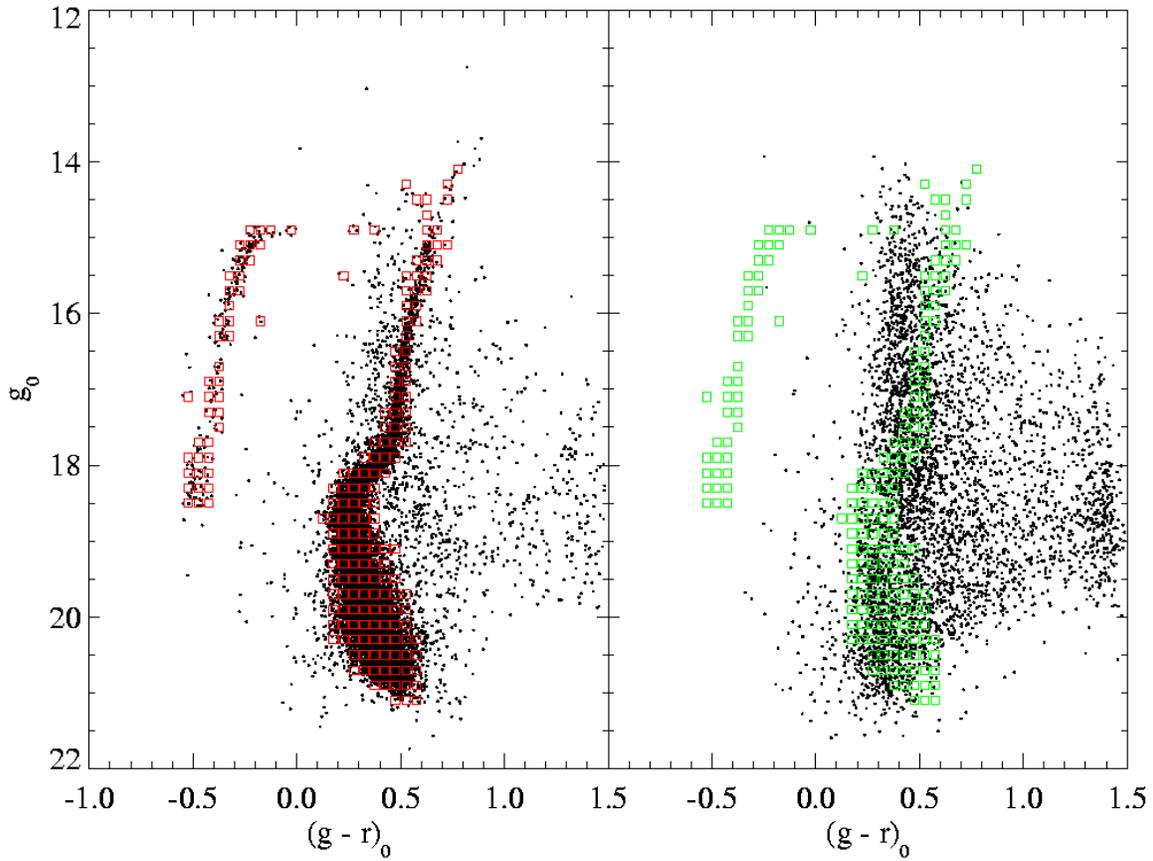}
\caption{Color-Magnitude Diagrams of the M~13 stars inside the tidal
radius (left panel), and inside the field region (right panel),
shown as black dots. The selected sub-grids from the $S/N$ cut are
shown as red squares in the left panel and green squares in the
right panel. These selected sub-grids are used in the analysis.}

\end{figure}
\clearpage

%% Figure 3
\begin{figure}
\centering \plotone{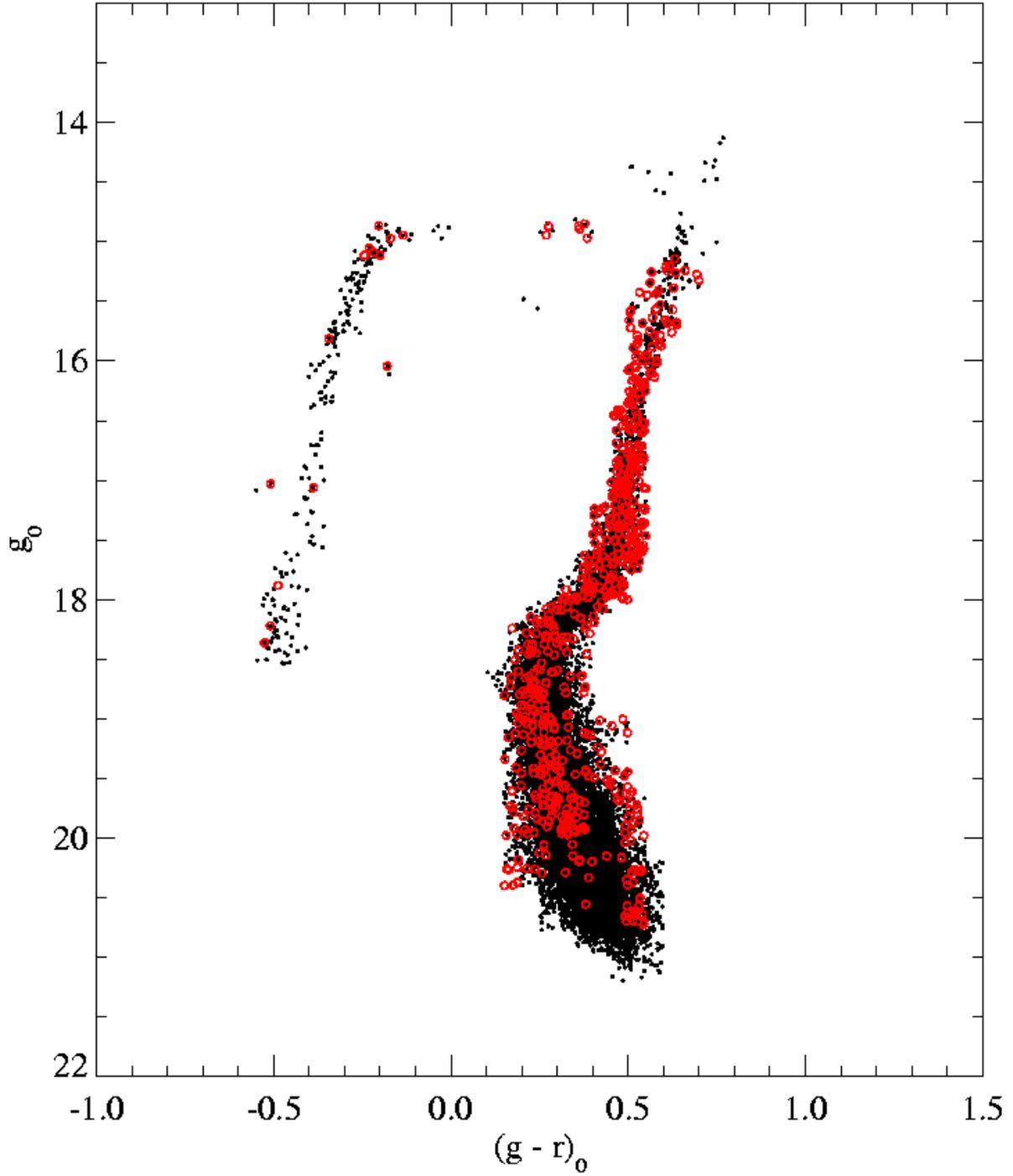}
%\plotone{f6.eps}
\caption{The M~13 Color-Magnitude Diagram based on the likely member
stars (black dots) selected from the photometric sample. The likely
members identified from the spectroscopic sample are indicated with
red circles.}
\end{figure}
\clearpage

%% Figure 4
\begin{figure}
\centering \plotone{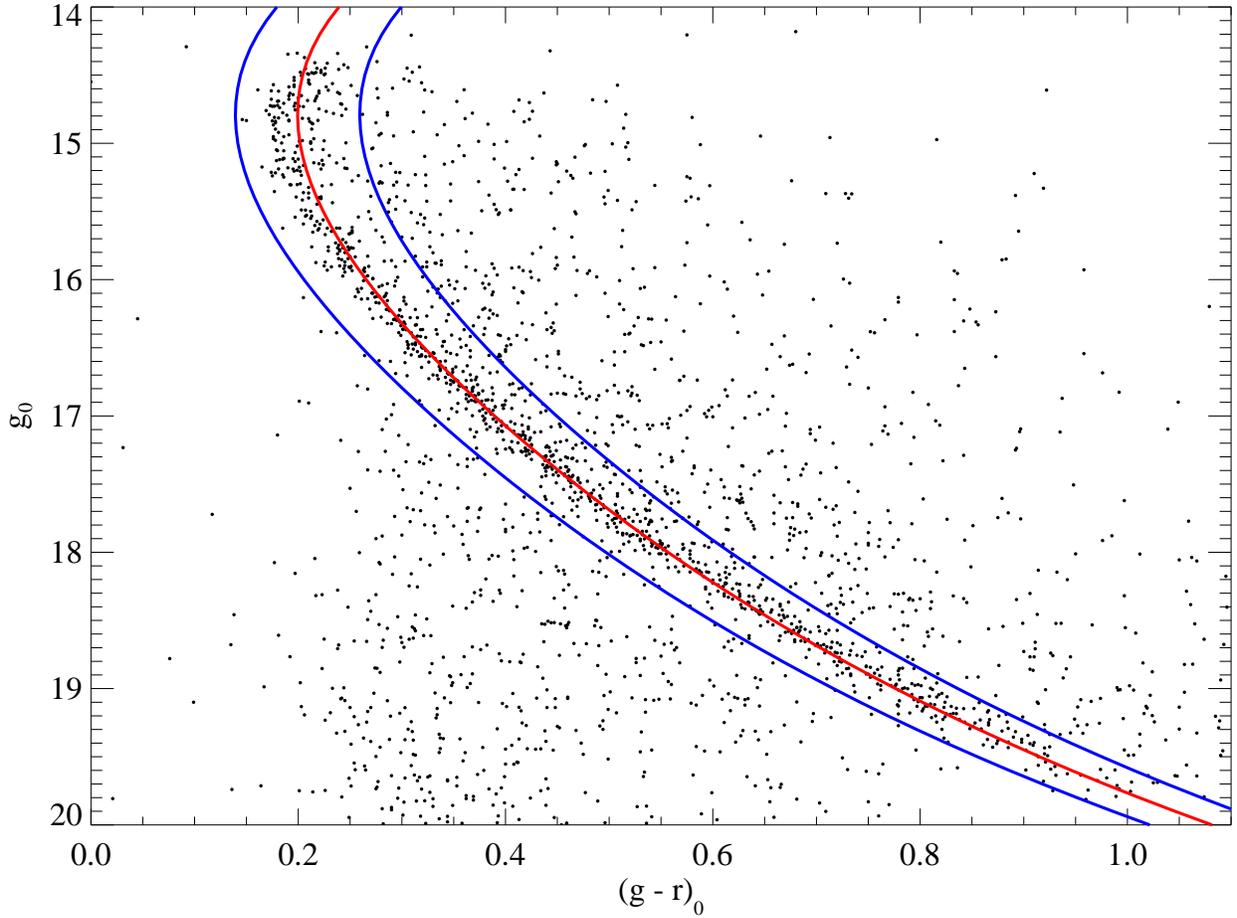}
%\plotone{n2420cmdsel}
\caption{Color-Magnitude Diagram of the NGC~2420 field. The red
line is the fiducial line obtained by application of a robust
fourth-order polynomial fit. The stars inside the blue lines
(fiducial $\pm$ 0.06 mag in $(g-r)_0$) are regarded as likely member
stars from the photometric sample.}
\end{figure}
\clearpage

%% Figure 5
\begin{figure}
\centering
\includegraphics[angle=90,scale=0.60]{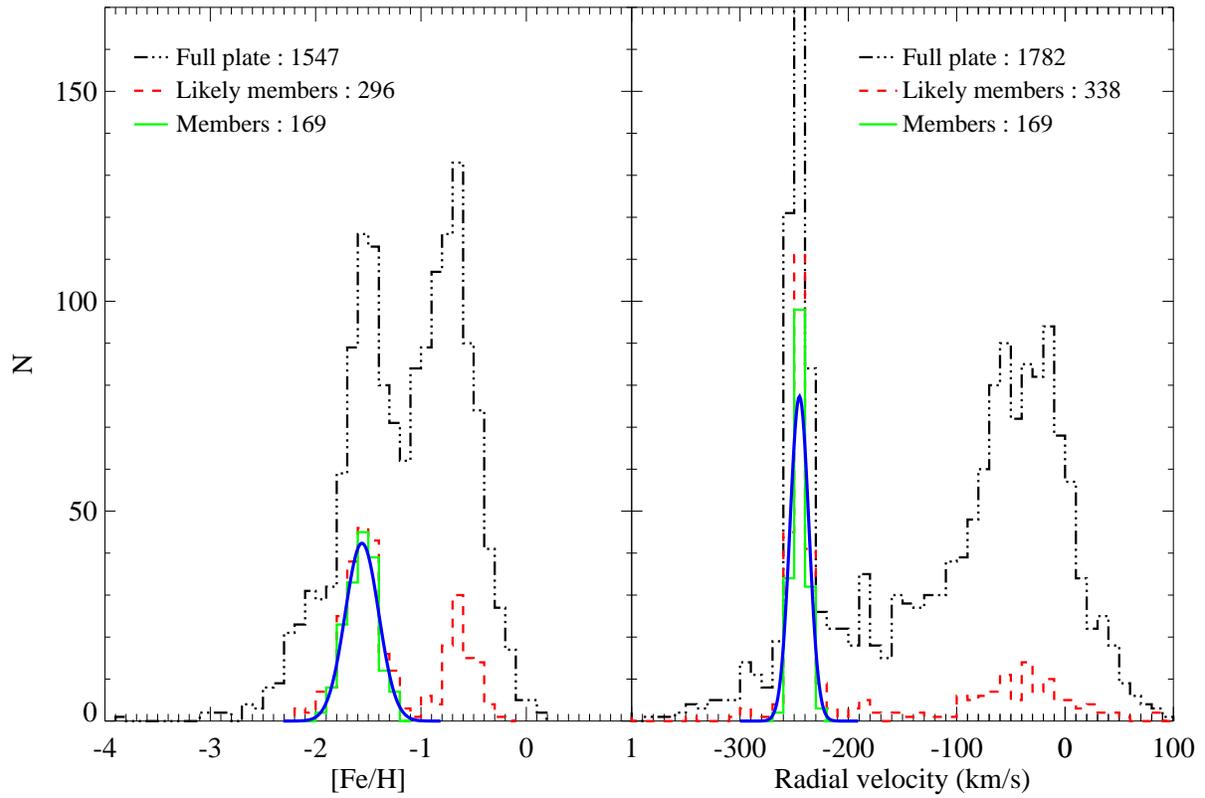}
\caption{[Fe/H] and radial velocity distributions for stars in the
direction of M~13. Gaussian fits (blue solid curves) to the
distribution of the selected true members, shown in the green
histograms, are over-plotted. }
\end{figure}
\clearpage

%% Figure 6
\begin{figure}
\centering
\includegraphics[angle=90,scale=0.60]{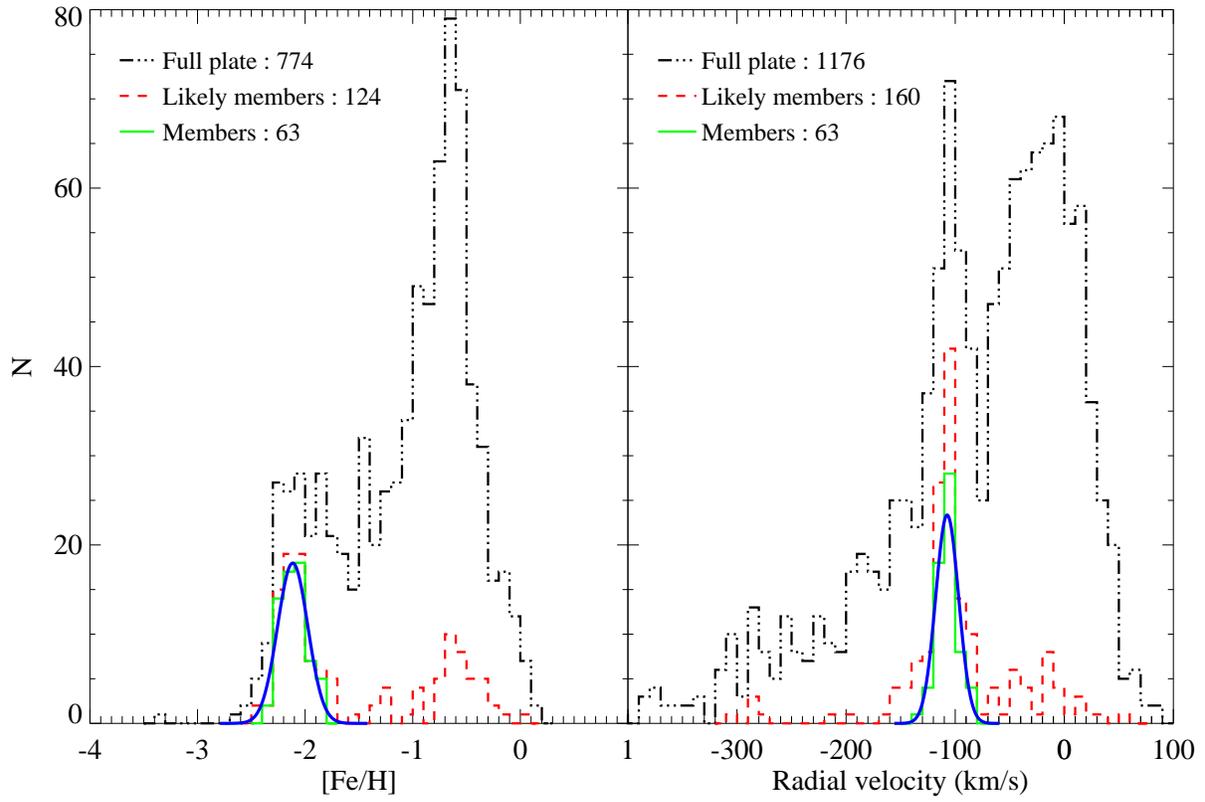}
\caption{Same as Fig. 5 but for M~15.}
%\caption{[Fe/H] and radial velocity
%distributions for stars in the direction of M~15. Gaussian fits
%(blue solid curves) to the distribution of the selected true
%members, shown in the green histograms, are over-plotted.}
\end{figure}
\clearpage

%% Figure 7
\begin{figure}
\centering
\includegraphics[angle=90,scale=0.60]{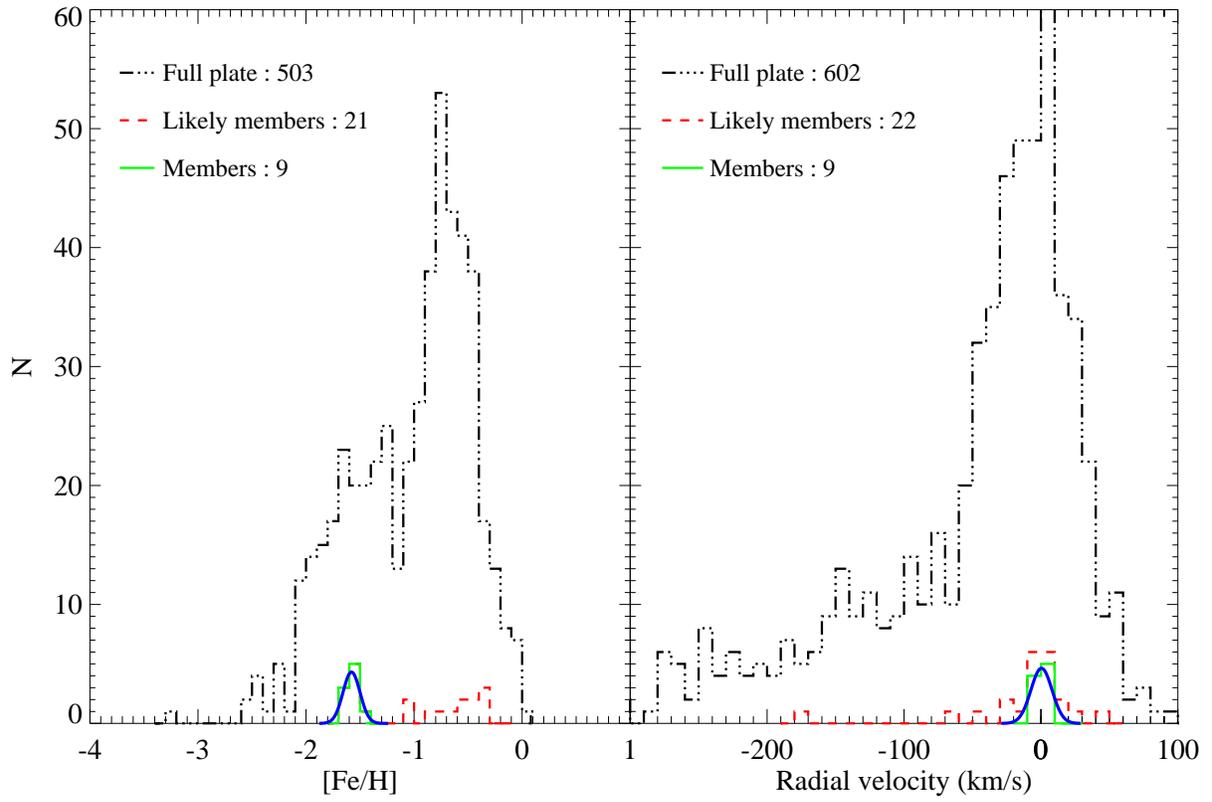}
\caption{Same as Fig. 5 but for M~2.}
%\caption{[Fe/H] and radial velocity
%distributions for stars in the direction of M~2. Gaussian fits (blue
%solid curves) to the distribution of the selected true members,
%shown in the green histograms, are over-plotted.}
\end{figure}
\clearpage

%% Figure 8
\begin{figure}
\centering
\includegraphics[angle=90,scale=0.60]{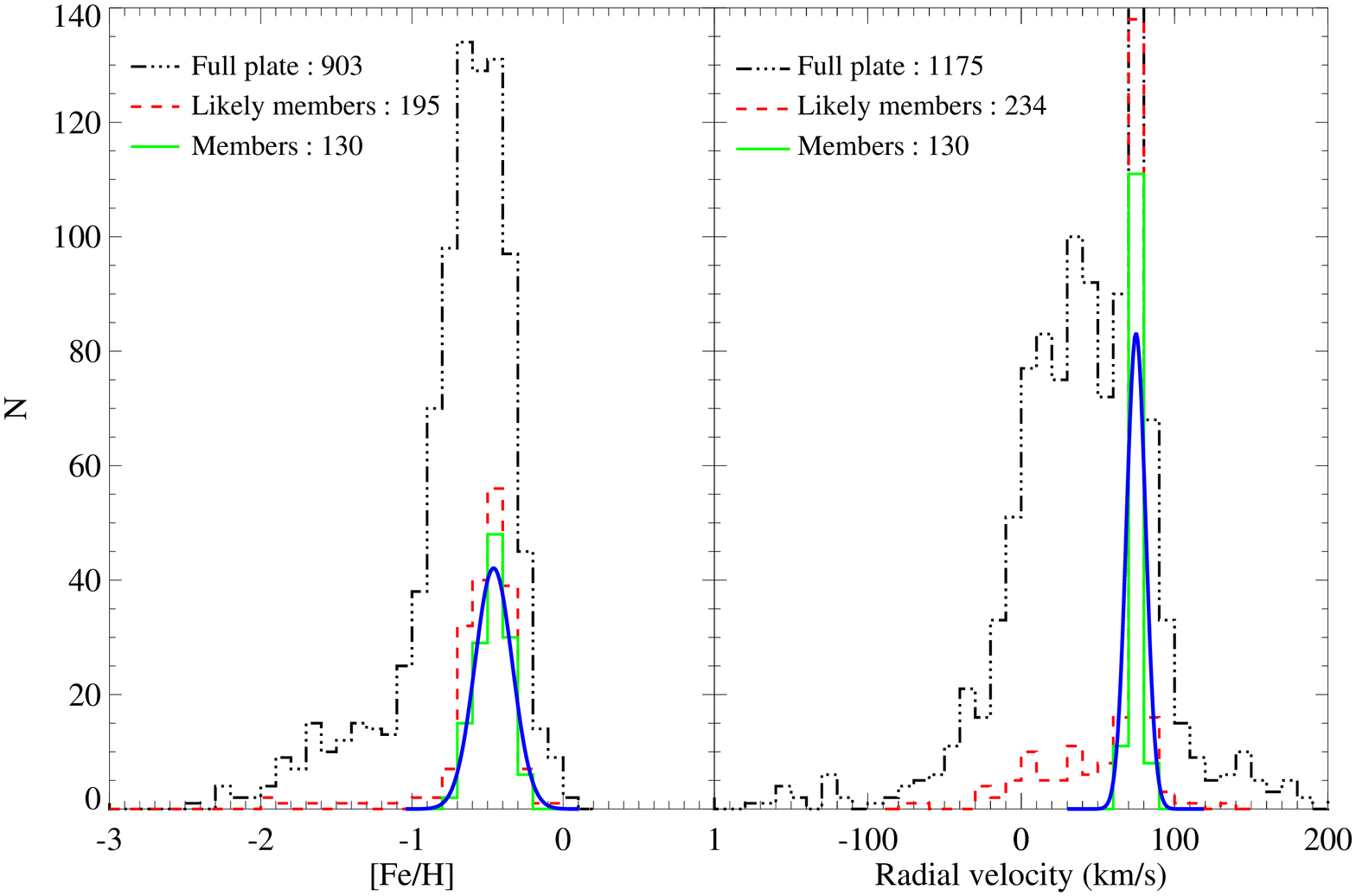}
\caption{Same as Fig. 5 but for NGC~2420.}
%\caption{[Fe/H] and radial velocity
%distributions for stars in the direction of NGC~2420. Gaussian fits
%(blue solid curves) to the distribution of the selected true
%members, shown in the green histograms, are over-plotted.}
\end{figure}
\clearpage

%% Figure 9
\begin{figure}
\centering
\includegraphics[angle=90,scale=0.60]{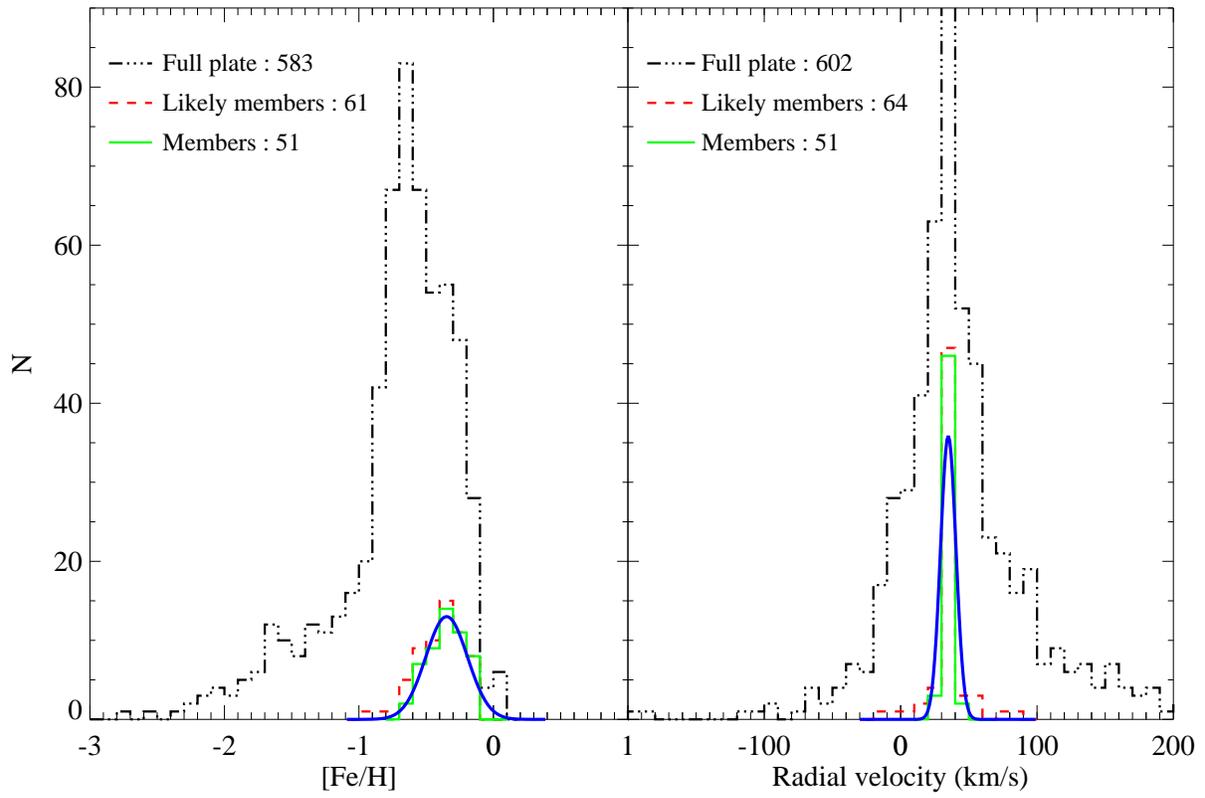}
\caption{Same as Fig. 5 but for M~67.}
%\caption{[Fe/H] and radial velocity
%distributions for stars in the direction of M~67. Gaussian fits
%(blue solid curves) to the distribution of the selected true
%members, shown in the green histograms, are over-plotted.}
\end{figure}

%% Figure 10
\begin{figure}
\centering
\includegraphics[scale=0.8]{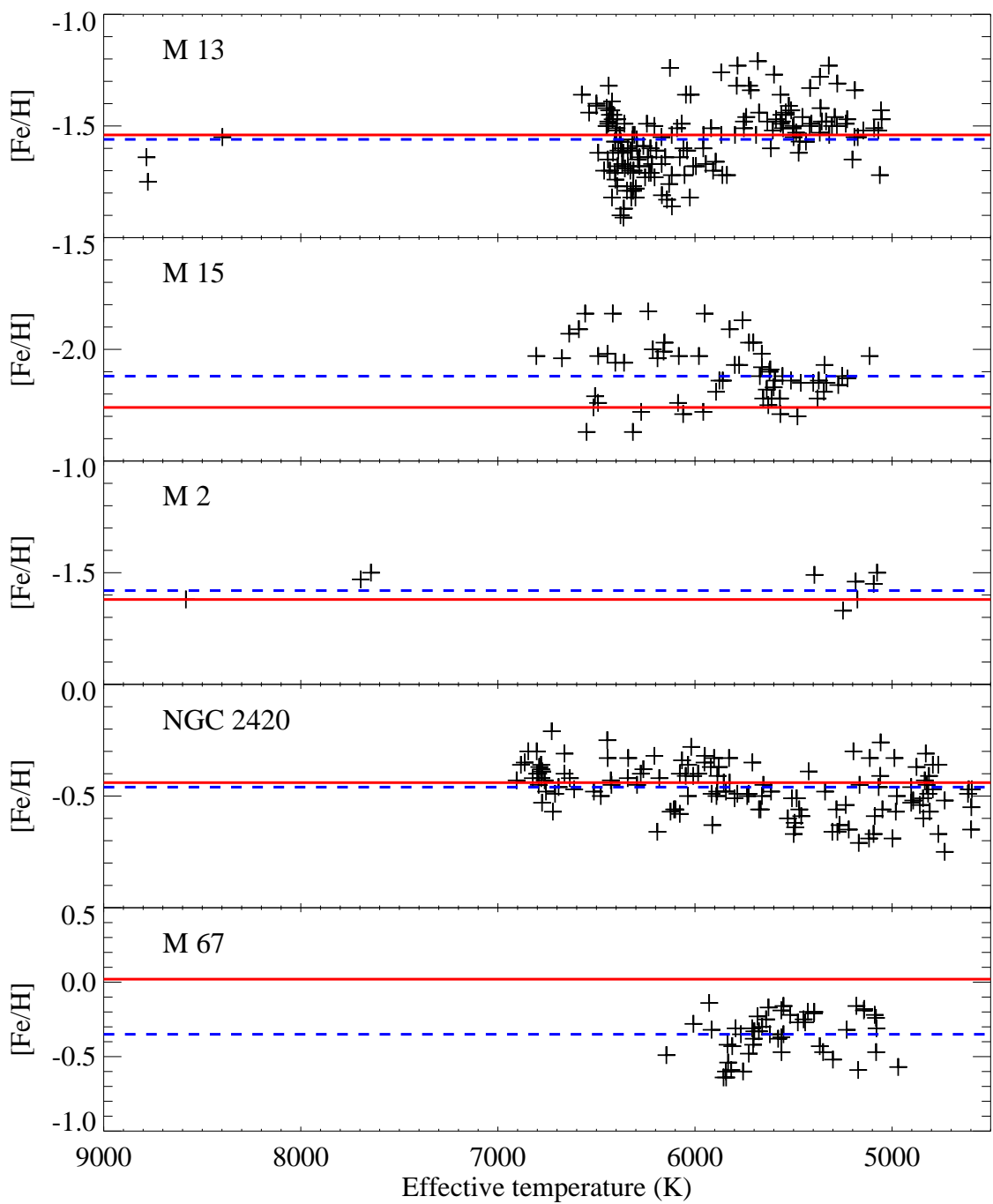}
\caption{Distribution of [Fe/H] as a function of effective
temperature for selected true member stars of M~13, M~15, M~2, NGC
2420, and M 67. The mean [Fe/H] determined for each cluster based on
these estimates is shown with the blue dashed line; the red solid
line represents the adopted literature value in each panel.}
\end{figure}
\clearpage

%% Figure 11
\begin{figure}
\centering
\includegraphics[angle=90,scale=0.60]{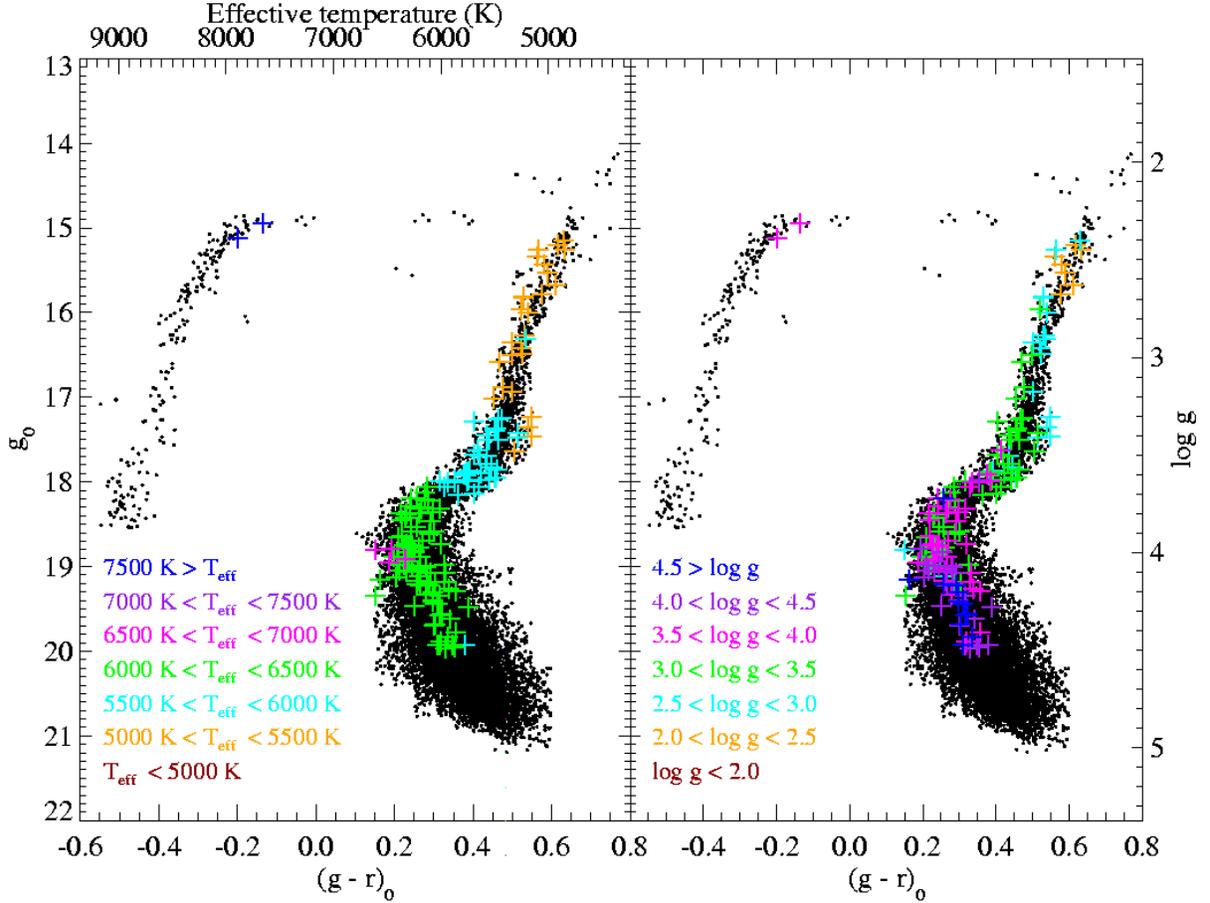}
\caption{Temperature and gravity distributions of the selected true
member stars for M~13. Each color represents a temperature range of
width 500~K, and a surface gravity range of 0.5 dex, respectively.
The temperature scales on the top of the left hand panel come from a
linear relation between $(g-r)_0$ color and $T_{\rm eff}$ by
performing a least squares fit to the theoretical models of Girardi
et al. (2004). A similar procedure is applied for transforming the
$g_0$ magnitude to a theoretical log $g$ scale on the ordinate in
the right-hand panel.}
\end{figure}
\clearpage

%% Figure 12
\begin{figure}
\includegraphics[angle=90,scale=0.60]{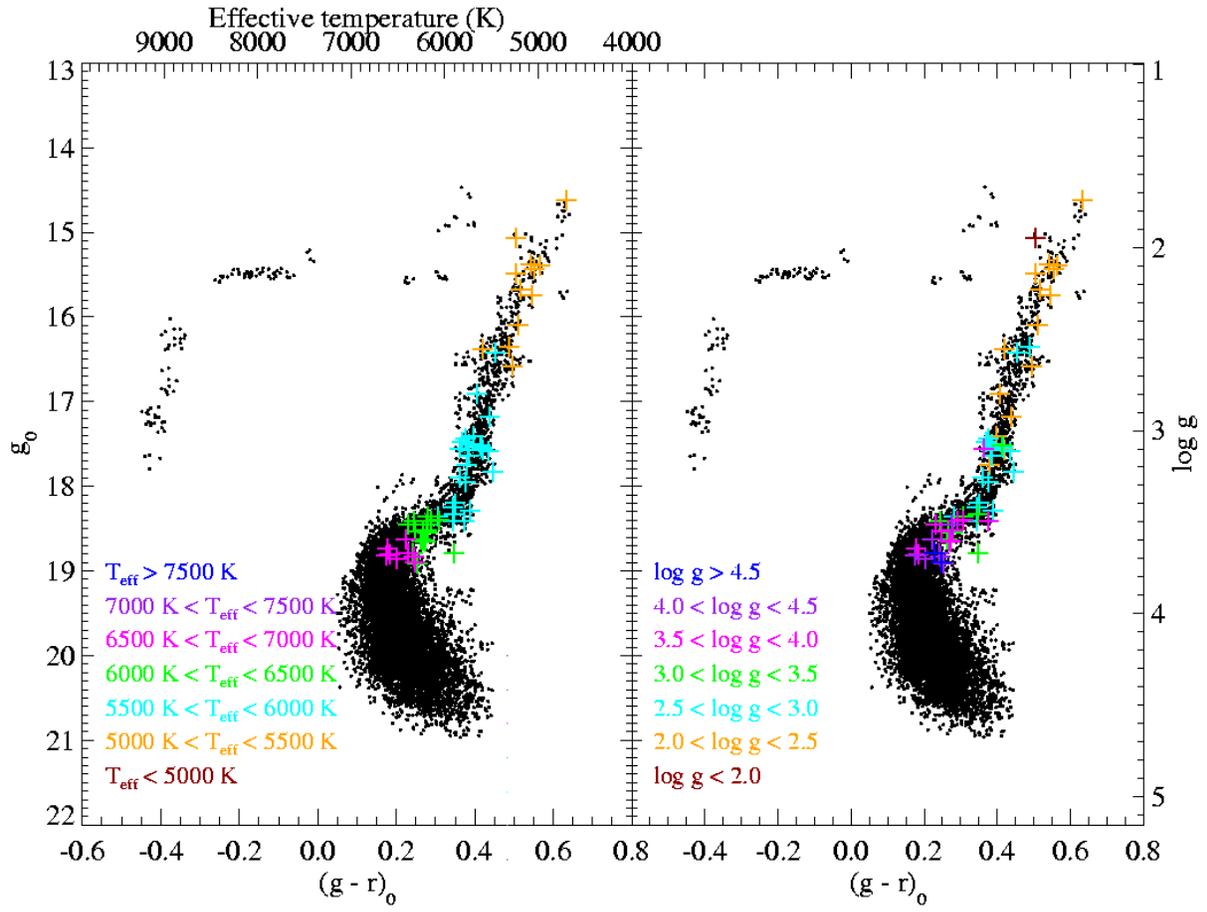}
\caption{Same as Fig. 11 but for M~15.}
%\caption{Temperature and gravity
%distributions of the selected true member stars for M~15. Each color
%represents a temperature range of width 500~K, and a surface gravity
%range of 0.5 dex, respectively. The temperature scales on the top of
%the left hand panel come from a linear relation between $(g-r)_0$
%color and $T_{\rm eff}$ by performing a least squares fit to the
%theoretical models of Girardi et al. (2004). A similar procedure is
%applied for transforming the $g_0$ magnitude to a theoretical log
%$g$ scale on the ordinate in the right-hand panel.}
\end{figure}
\clearpage

%% Figure 13
\begin{figure}
\centering
\includegraphics[angle=90,scale=0.60]{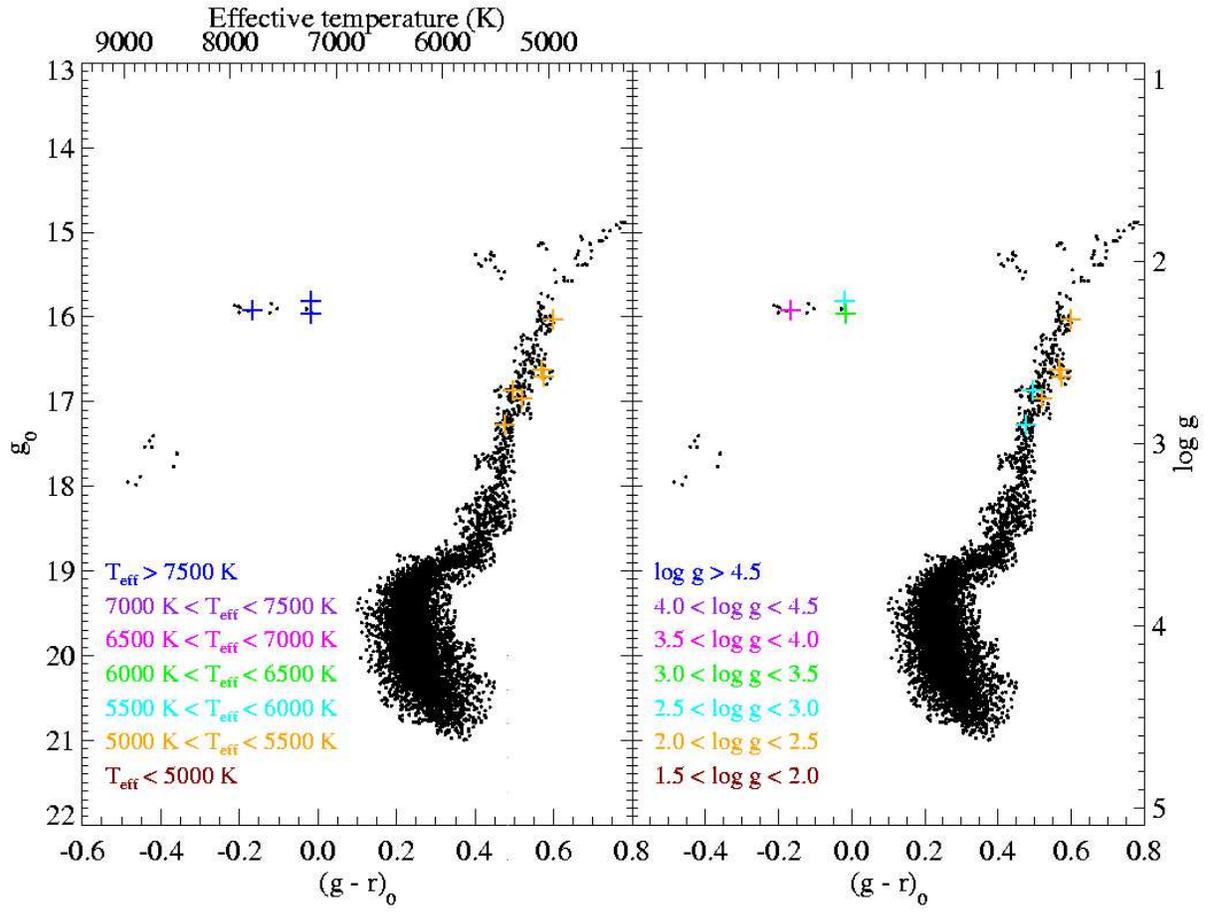}
\caption{Same as Fig. 11 but for M~2.}
%\caption{Temperature and gravity
%distributions of the selected true member stars for M~2. Each color
%represents a temperature range of width 500~K, and a surface gravity
%range of 0.5 dex, respectively. The temperature scales on the top of
%the left hand panel come from a linear relation between $(g-r)_0$
%color and $T_{\rm eff}$ by performing a least squares fit to the
%theoretical models of Girardi et al. (2004). A similar procedure is
%applied for transforming the $g_0$ magnitude to a theoretical log
%$g$ scale on the ordinate in the right-hand panel.}
\end{figure}
\clearpage

%% Figure 14
\begin{figure}
\centering
\includegraphics[angle=90,scale=0.60]{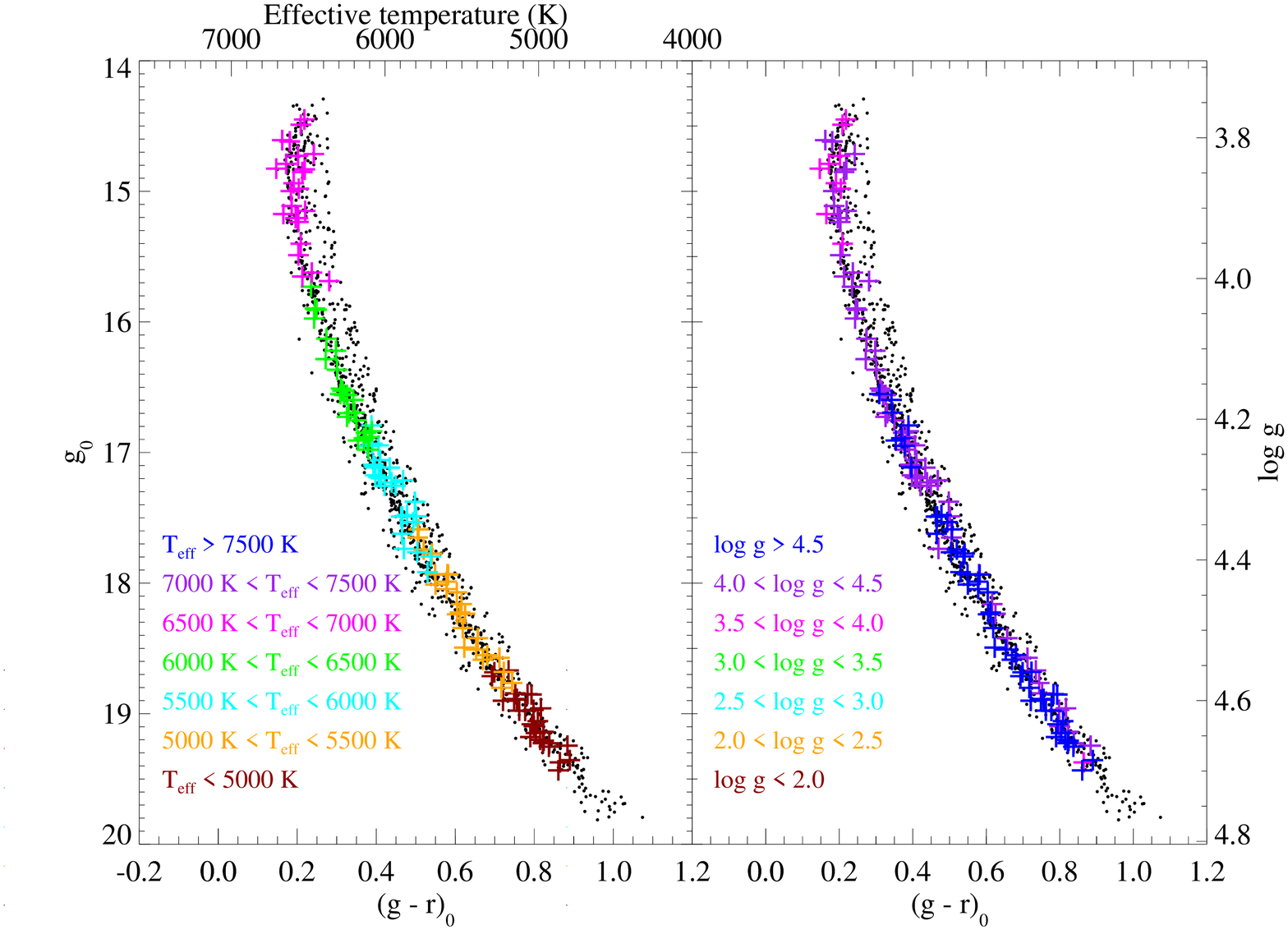}
\caption{Same as Fig. 11 but for NGC~2420.}
%\caption{Temperature and gravity distributions of the selected true
%member stars for NGC~2420. Each color represents a temperature range
%of width 500~K, and a surface gravity range of 0.5 dex,
%respectively. The temperature scales on the top of the left hand
%panel come from a linear relation between $(g-r)_0$ color and
%$T_{\rm eff}$ by performing a least squares polynomial fit to the
%theoretical models of Girardi et al. (2004). A similar procedure is
%applied for transforming the $g_0$ magnitude to a theoretical log
%$g$ scale on the ordinate in the right-hand panel.}
\end{figure}
\clearpage

% Figure 15
\begin{figure}
\centering
\includegraphics[angle=90,scale=0.60]{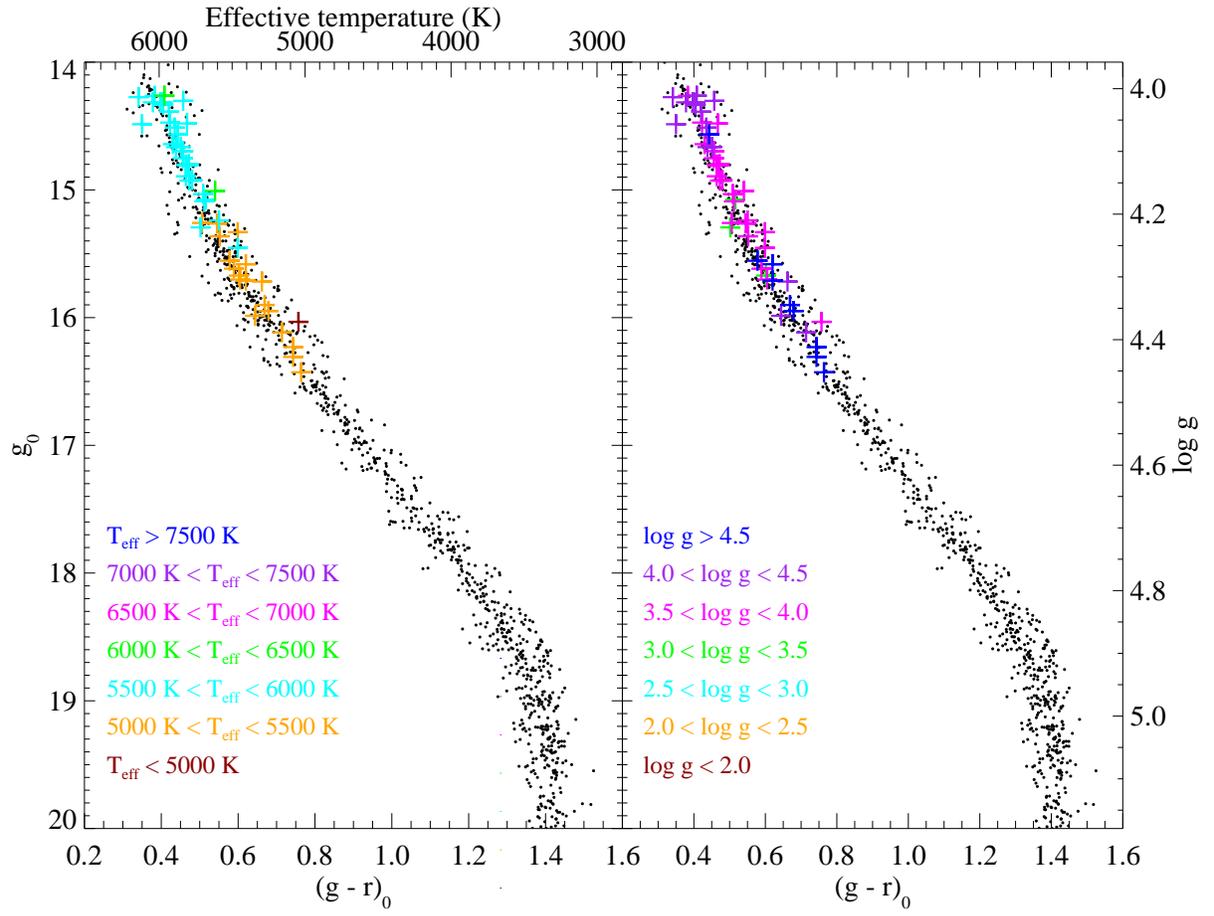}
\caption{Same as Fig. 11 but for M~67.}
%\caption{Temperature and gravity
%distributions of the selected true member stars for M~67. Each color
%represents a temperature range of width 500~K, and a surface gravity
%range of 0.5 dex, respectively. The temperature scales on the top of
%the left hand panel come from a linear relation between $(g-r)_0$
%color and $T_{\rm eff}$ by performing a least squares fit to the
%theoretical models of Girardi et al. (2004). A similar procedure is
%applied for transforming the $g_0$ magnitude to a theoretical log
%$g$ scale on the ordinate in the right-hand panel.}
\end{figure}
\clearpage

%% Table 1
%%\documentclass[12pt,preprint]{aastex}
%%\pagestyle{empty}
\begin{deluxetable}{lccccc}
\tablecolumns{4} \tablewidth{0pc} \tablecaption{Properties of the
Clusters} \tablehead{ \colhead{} & \colhead{M 15} &
\colhead{M 2} & \colhead{M 13 } & \colhead{NGC 2420} & \colhead{M 67} \\
\colhead{} & \colhead{(NGC 7078)} & \colhead{(NGC 7089)} &
\colhead{(NGC 6205)} & \colhead{} & \colhead{(NGC 2682)}}
\startdata \rm RA (J2000) & 21:29:58.3 & 21:33:29.3 & 16:41:41.5 & 07:38:23 & 08:51:18 \\
\rm DEC (J2000) & $+$12:10:01 & $-$00:49:23 & $+$36:27:37 & $+$21:34:24 & $+$11:48:00 \\
$(l, b)$ & (65.0, $-$27.3) & (58.4, $-$35.8)& (59.0, $+$40.9) & (198.1, $+$19.6) & (215.7, $+$31.9) \\
\rm [Fe/H] & $-$2.26 & $-$1.62 & $-$1.54 & $-$0.44 & $+$0.02\\
\rm m - M & 15.37 & 15.49 & 14.48 & 12.54 & 9.97 \\
\rm V$_{r}(\rm km~s^{-1}$) & $-$107.0 & $-$5.3 & $-$245.6 & 74.0 & 32.9\\
E$(B-V)$ & 0.1 & 0.06 & 0.02 & 0.03 & 0.06 \\
r$_{t}$ & 21.50 &21.45 & 25.18 & 5\tablenotemark{1} & 25\tablenotemark{1} \\
\enddata
\tablenotetext{1}{This case is the apparent diameter of the cluster
in arc minutes from Dias et al. (2002).} \tablecomments{The parameter r$_{t}$ is
the tidal radius in arc minutes. The listed distance modulus, m$-$M, is corrected for
extinction. The parameters for the globular clusters, M~13, M~15, and M~2
come from Harris (1996); those for the open clusters NGC~2420 and M~67 are
from WEBDA (http://www.univie.ac.at/webda/). The radial velocity
(V$_{r}$) listed for NGC~2420 and M~67 is the average of three literature
values listed in WEBDA (see text). The [Fe/H] of NGC~2420 and M~67 is based on high-
resolution spectroscopy described by Gratton (2000).}
\end{deluxetable}
%%\end{document}

\clearpage

%% Table 2 begins
%%\documentclass[12pt,preprint]{aastex}
%%\pagestyle{empty}
\begin{deluxetable}{lccrrccrrccrc}
\renewcommand{\tabcolsep}{2pt}
\tabletypesize{\scriptsize} \tablecolumns{13} \tablewidth{0pc}
\tablecaption{Derived Radial Velocities and Metallicities of the
Clusters} \tablehead{ \colhead{Cluster} & \colhead{[Fe/H]$_{l}$} &
\colhead{[Fe/H]$_{u}$} & \colhead{RV$_{l}$} & \colhead{RV$_{u}$} &
\colhead{$<$[Fe/H]$>$} & \colhead{$\sigma$ ([Fe/H])} &
\colhead{$<$RV$>$} & \colhead{$\sigma$ (RV)} & \colhead{Mem$_{\rm
tot}$} & \colhead{[Fe/H]$_{\rm H}$} &
\colhead{RV$_{\rm H}$} & \colhead{[Fe/H]$_{\rm HR}$}   \\
\colhead{ } & \colhead{dex} & \colhead{dex} & \colhead{km s$^{-1}$}
& \colhead{km s$^{-1}$} & \colhead{dex} & \colhead{dex} &
\colhead{km s$^{-1}$} & \colhead{km s$^{-1}$} & \colhead{} &
\colhead{dex)} & \colhead{km s$^{-1}$} &
\colhead{dex}\\
\colhead{(1)} & \colhead{(2)} & \colhead{(3)} & \colhead{(4)} &
\colhead{(5)} & \colhead{(6)} & \colhead{(7)} & \colhead{(8)} &
\colhead{(9)} & \colhead{(10)} & \colhead{(11)} & \colhead{(12)} &
\colhead{(13)} }

\startdata
\rm M 13     & $-$1.91 & $-$1.20 & $-$264.0 & $-$226.1 & $-$1.56 & 0.16  & $-$245.1 &  8.7  & 169 & $-$1.54 & $-$245.6 & $-$1.63 \\
\rm M 15     & $-$2.42 & $-$1.80 & $-$132.8 &  $-$81.6 & $-$2.12 & 0.14  & $-$107.4 & 10.5  &  63 & $-$2.26 & $-$107.0 & $-$2.42 \\
\rm M 2      & $-$1.74 & $-$1.40 & $-$24.5  &  $+$25.3 & $-$1.58 & 0.08  & $+$0.4   &  7.7  &  9  & $-$1.62 & $-$  5.3 & $-$1.56 \\
\rm NGC 2420 & $-$0.77 & $-$0.19 & $+$60.5  &  $+$89.3 & $-$0.46 & 0.12  & $+$74.8  &  6.2  & 130 & $\cdot\cdot\cdot\cdot$ & $+$  74.0\tablenotemark{1} & $-$0.44 \\
\rm M 67     & $-$0.73 & $-$0.02 & $+$21.1  &  $+$48.6 & $-$0.35 & 0.16  & $+$34.9   & 5.6  &  51 & $\cdot\cdot\cdot\cdot$ & $+$ 32.9\tablenotemark{1} & $+$0.02 \\
\enddata
\tablenotetext{1}{This value is an average from the literature
values. See \S 4.2 for more detail.} \tablecomments{The subscripts
$u$ and $l$ on [Fe/H] and RV are the upper and lower limits for the
metallicity and radial velocity cuts used for choosing true members,
as described in the text. The parameters for the globular clusters
M~13, M~15, and M~2 come from Harris (1996); those for the open
clusters NGC~2420 and M~67 are from WEBDA (http:
//www.univie.ac.at/webda/). Values of [Fe/H]$_{\rm HR}$, which are based on
high-resolution spectroscopy, are from Kraft $\&$ Ivans (2003; M~13
and ~15) from Ivans (private communication; M~2), and from Gratton
(2000; NGC~2420 and M~67). The column labeled Mem$_{\rm tot}$ lists
the total number of true members assigned to each cluster.}
\end{deluxetable}

%%\end{document}

\clearpage

%% Table 3 begins
%%\documentclass[12pt,preprint]{aastex}
%%\pagestyle{empty}
%%\begin{document}
\begin{deluxetable}{lcrccc}
\tablecolumns{5} \tablewidth{0pc} \tablecaption{Mean and 1$\sigma$
Values of Residuals of $T_{\rm eff}$ and $\log g$ Between SSPP and
Model Fits.} \tablehead{ \colhead{} & \colhead{} &
\multicolumn{2}{c}{$T_{\rm eff}$} & \multicolumn{2}{c}{$\log g$} \\
\colhead{Cluster} & \colhead{N} & \colhead{$<\Delta>$} &
\colhead{$\sigma$} &
\colhead{$<\Delta>$} & \colhead{$\sigma$} \\
\colhead{} & \colhead{} & \colhead{(K)} & \colhead{(K)} &
\colhead{(dex)} & \colhead{(dex)}} \startdata
\rm M 13     & 169 & $+$51  & 125 & $-$0.07 &  0.44 \\
\rm M 15     & 63  & $+$110 & 94  & $-$0.17 &  0.40 \\
\rm M 2      & 9   & $+$190 & 293 & $+$0.19 &  0.55 \\
\rm NGC 2420 & 130 & $+$12  & 105 & $+$0.15 &  0.42 \\
\rm M 67     & 51  & $-$119 & 104 & $-$0.27 &  0.43 \\
\enddata
\tablecomments{These values are 5$\sigma$ clipped averages and
standard deviations.}
\end{deluxetable}

%%\end{document}

\clearpage

%% Table 3
%%\documentclass[12pt,preprint]{aastex}
%%\pagestyle{empty}
\begin{deluxetable}{cccrccccccccccccccccccc}
\tabletypesize{\scriptsize} \tablenum{4} \tablecolumns{22}
\tablewidth{0in} \rotate
\renewcommand{\tabcolsep}{2pt} \tablecaption{Properties of Selected Member Stars
of M~13} \tablehead{\colhead{spSpec name} & \colhead{RA} &
\colhead{DEC} & \colhead{RV} & \colhead{$\sigma_{\rm RV}$} &
\colhead{T$_{\rm eff}$} & \colhead{$\sigma_{\rm T_{\rm eff}}$} &
\colhead{log $g$} & \colhead{$\sigma_{\rm log \it g}$} & \colhead
{[Fe/H]} & \colhead{$\sigma_{\rm [Fe/H]}$} & \colhead{$u$} &
\colhead{$\sigma_{u}$} & \colhead{$g$} & \colhead{$\sigma_{g}$} &
\colhead{$r$} & \colhead{$\sigma_{r}$} & \colhead{$i$} &
\colhead{$\sigma_{i}$} & \colhead{$z$} & \colhead{$\sigma_{z}$} &
\colhead{$\rm <S/N>$} \\

\colhead{} & \colhead{(degree)} & \colhead{(degree)} & \colhead{(km
s$^{-1}$)} & \colhead{(km s$^{-1}$)} & \colhead{(K)} & \colhead{(K)}
& \colhead{(dex)} & \colhead{(dex)} & \colhead {(dex)} &
\colhead{(dex)} & \colhead{} & \colhead{} & \colhead{} & \colhead{}
& \colhead{} & \colhead{} & \colhead{} & \colhead{} & \colhead{} &
\colhead{} & \colhead{}}

\startdata
      53521$-$2174$-$054  &   250.8113861  &     36.380367  &   $-$240.4  &      5.2  &     5605  &        2  &    2.98  &     0.22  &     $-$1.52  &     0.09  &   19.00  &    0.03  &   17.83  &    0.01  &   17.39  &    0.02  &   17.23  &    0.02  &   17.17  &    0.02  &   17.1  \\
      53521$-$2174$-$082  &   250.6566925  &     36.292824  &   $-$244.1  &      2.0  &     5146  &       44  &    2.35  &     0.23  &     $-$1.52  &     0.06  &   16.88  &    0.02  &   15.43  &    0.01  &   14.85  &    0.02  &   14.63  &    0.02  &   14.50  &    0.02  &   51.2  \\
      53521$-$2174$-$087  &   250.5986176  &     36.141987  &   $-$236.8  &      3.5  &     5500  &       31  &    3.11  &     0.21  &     $-$1.55  &     0.06  &   18.42  &    0.02  &   17.26  &    0.02  &   16.79  &    0.02  &   16.62  &    0.02  &   16.54  &    0.02  &   24.3  \\
      53521$-$2174$-$094  &   250.6187286  &     36.193378  &   $-$241.2  &      3.4  &     5553  &       25  &    3.14  &     0.27  &     $-$1.51  &     0.02  &   18.42  &    0.02  &   17.25  &    0.02  &   16.78  &    0.02  &   16.56  &    0.02  &   16.47  &    0.02  &   24.9  \\
      53521$-$2174$-$126  &   250.4946899  &     36.287205  &   $-$235.8  &      3.8  &     5487  &       60  &    3.02  &     0.12  &     $-$1.53  &     0.03  &   18.25  &    0.03  &   17.01  &    0.01  &   16.56  &    0.01  &   16.31  &    0.03  &   16.28  &    0.02  &   27.1  \\
      53521$-$2174$-$128  &   250.4741516  &     36.309807  &   $-$252.5  &      4.3  &     5524  &       77  &    3.34  &     0.24  &     $-$1.41  &     0.06  &   18.64  &    0.03  &   17.50  &    0.01  &   17.06  &    0.01  &   16.77  &    0.03  &   16.75  &    0.02  &   19.2  \\
      53521$-$2174$-$133  &   250.5125122  &     36.321091  &   $-$245.7  &      3.4  &     5418  &       64  &    3.16  &     0.11  &     $-$1.50  &     0.08  &   18.10  &    0.03  &   16.88  &    0.01  &   16.40  &    0.01  &   16.14  &    0.03  &   16.11  &    0.04  &   31.8  \\
\enddata
\tablecomments{The column labeled spSpec name is constructed from
the Modified Julian Date (5 digits), the spectroscopic plug plate
number (four digits), and the fiber used (three digits). The error
in the listed RV is computed from ELODIE template matching. The
atmospheric parameter estimates ($T_{\rm eff}$, log $g$, and [Fe/H])
are the adopted values from the SSPP (Lee et al. 2007a); the quoted
internal error is the listed uncertainty in each value. The
photometric errors in each band are based on the PHOTO pipeline
(Lupton et al. 2001). In the last column, $<$S/N$>$ is the average
signal-to-noise  per pixel of the spectra, calculated over the range
3850$-$6000\,{\AA}.}
\end{deluxetable}
%%\end{document}

%\input{m13tab.tex}
\clearpage

%% Table 4
%%\documentclass[12pt,preprint]{aastex}
%%\pagestyle{empty}
\begin{deluxetable}{cccrccccccccccccccccccc}
\tabletypesize{\scriptsize} \tablenum{5} \tablecolumns{22}
\tablewidth{0in} \rotate
\renewcommand{\tabcolsep}{2pt} \tablecaption{Properties of Selected Member Stars
of M~15} \tablehead{\colhead{spSpec name} & \colhead{RA} & \colhead{DEC}
& \colhead{RV} & \colhead{$\sigma_{\rm RV}$} & \colhead{T$_{\rm
eff}$} &
 \colhead{$\sigma_{\rm T_{\rm eff}}$} & \colhead{log $g$} &
 \colhead{$\sigma_{\rm log \it g}$} & \colhead {[Fe/H]} &
 \colhead{$\sigma_{\rm [Fe/H]}$} & \colhead{$u$} &
 \colhead{$\sigma_{u}$} & \colhead{$g$} & \colhead{$\sigma_{g}$} &
 \colhead{$r$} & \colhead{$\sigma_{r}$} & \colhead{$i$} &
 \colhead{$\sigma_{i}$} & \colhead{$z$} & \colhead{$\sigma_{z}$} &
 \colhead{$\rm <S/N>$} \\

\colhead{} & \colhead{(degree)} & \colhead{(degree)} & \colhead{(km
s$^{-1}$)} & \colhead{(km s$^{-1}$)} & \colhead{(K)} & \colhead{(K)}
& \colhead{(dex)} & \colhead{(dex)} & \colhead {(dex)} &
\colhead{(dex)} & \colhead{} & \colhead{} & \colhead{} & \colhead{}
& \colhead{} & \colhead{} & \colhead{} & \colhead{} & \colhead{} &
\colhead{} & \colhead{}}

\startdata
      53289$-$1960$-$401  &   322.4521790  &     12.338844  &   $-$107.5  &      2.8  &     5372  &       18  &    2.36  &     0.36  &    $-$2.14  &     0.05  &    17.24  &    0.02  &   16.09  &    0.01  &   15.58  &    0.01  &   15.35  &    0.01  &   15.28  &    0.01  &   48.2  \\
      53289$-$1960$-$402  &   322.4679565  &     12.327691  &   $-$107.1  &      2.4  &     5226  &       20  &    2.10  &     0.31  &    $-$2.13  &     0.05  &    16.71  &    0.02  &   15.39  &    0.01  &   14.83  &    0.01  &   14.56  &    0.01  &   14.48  &    0.01  &   54.3  \\
      53289$-$1960$-$406  &   322.4168396  &     12.266688  &   $-$100.7  &      2.1  &     5253  &       18  &    2.09  &     0.26  &    $-$2.12  &     0.04  &    16.69  &    0.02  &   15.38  &    0.01  &   14.83  &    0.01  &   14.57  &    0.01  &   14.48  &    0.01  &   58.2  \\
      53289$-$1960$-$411  &   322.4046326  &     12.227963  &   $-$107.9  &      3.8  &     5480  &        5  &    2.34  &     0.13  &    $-$2.30  &     0.05  &    17.48  &    0.02  &   16.38  &    0.01  &   15.96  &    0.01  &   15.71  &    0.01  &   15.60  &    0.01  &   41.2  \\
      53289$-$1960$-$413  &   322.4143066  &     12.305798  &   $-$104.5  &      3.1  &     5346  &       36  &    2.11  &     0.27  &    $-$2.19  &     0.04  &    16.89  &    0.02  &   15.67  &    0.01  &   15.16  &    0.01  &   14.93  &    0.01  &   14.86  &    0.01  &   49.9  \\
      53289$-$1960$-$419  &   322.4579773  &     12.303373  &   $-$110.3  &      2.1  &     5273  &       28  &    2.25  &     0.37  &    $-$2.16  &     0.05  &    16.70  &    0.02  &   15.44  &    0.01  &   14.88  &    0.01  &   14.62  &    0.01  &   14.54  &    0.01  &   58.3  \\
      53289$-$1960$-$441  &   322.5975342  &     12.257596  &   $-$118.9  &      7.3  &     5759  &      132  &    3.53  &     0.56  &    $-$1.87  &     0.18  &    18.49  &    0.03  &   17.55  &    0.01  &   17.19  &    0.01  &   17.00  &    0.01  &   16.94  &    0.02  &   18.7  \\
\enddata
\end{deluxetable}
%%\end{document}

%\input{m15tab.tex}
\clearpage

%% Table 5
%%\documentclass[12pt,preprint]{aastex}
%%\pagestyle{empty}
\begin{deluxetable}{cccrccccccccccccccccccc}
\centering \tabletypesize{\scriptsize} \tablenum{6}
\tablecolumns{22} \tablewidth{0in} \rotate
\renewcommand{\tabcolsep}{2pt} \tablecaption{Properties of Selected Member Stars
of M~2} \tablehead{\colhead{spSpec name} & \colhead{RA} & \colhead{DEC}
& \colhead{RV} & \colhead{$\sigma_{\rm RV}$} & \colhead{T$_{\rm
eff}$} &
 \colhead{$\sigma_{\rm T_{\rm eff}}$} & \colhead{log $g$} &
 \colhead{$\sigma_{\rm log \it g}$} & \colhead {[Fe/H]} &
 \colhead{$\sigma_{\rm [Fe/H]}$} & \colhead{$u$} &
 \colhead{$\sigma_{u}$} & \colhead{$g$} & \colhead{$\sigma_{g}$} &
 \colhead{$r$} & \colhead{$\sigma_{r}$} & \colhead{$i$} &
 \colhead{$\sigma_{i}$} & \colhead{$z$} & \colhead{$\sigma_{z}$} &
 \colhead{$\rm <S/N>$} \\
\colhead{} & \colhead{(degree)} & \colhead{(degree)} & \colhead{(km
s$^{-1}$)} & \colhead{(km s$^{-1}$)} & \colhead{(K)} & \colhead{(K)}
& \colhead{(dex)} & \colhead{(dex)} & \colhead {(dex)} &
\colhead{(dex)} & \colhead{} & \colhead{} & \colhead{} & \colhead{}
& \colhead{} & \colhead{} & \colhead{} & \colhead{} & \colhead{} &
\colhead{} & \colhead{}}

\startdata
      53299$-$1961$-$123  &   323.3084412  &     $-$1.018151  &      9.5  &      2.3  &      7644  &      176  &   2.57  &     0.20  &     $-$1.50  &     0.10  &   16.97  &    0.02  &   15.81  &    0.02  &   15.83  &    0.01  &   15.84  &    0.01  &   15.88  &    0.01  &   44.6  \\
      53299$-$1961$-$125  &   323.3046265  &     $-$0.900651  &     $-$4.4  &      1.6  &      5076  &       52  &   2.17  &     0.21  &     $-$1.50  &     0.07  &   17.48  &    0.02  &   16.03  &    0.02  &   15.43  &    0.01  &   15.21  &    0.01  &   15.06  &    0.01  &   46.8  \\
      53299$-$1961$-$131  &   323.4663696  &     $-$0.819519  &      2.2  &      1.9  &      5185  &       34  &   2.40  &     0.25  &     $-$1.54  &     0.08  &   18.03  &    0.02  &   16.71  &    0.01  &   16.13  &    0.01  &   15.88  &    0.01  &   15.81  &    0.01  &   37.0  \\
      53299$-$1961$-$140  &   323.3161316  &     $-$0.936875  &     $-$9.1  &      3.0  &      5394  &       64  &   2.85  &     0.14  &     $-$1.51  &     0.08  &   18.53  &    0.02  &   17.27  &    0.03  &   16.79  &    0.01  &   16.59  &    0.01  &   16.52  &    0.01  &   27.9  \\
      53299$-$1961$-$144  &   323.4824524  &     $-$0.780800  &      3.2  &      2.8  &      5249  &       68  &   2.40  &     0.33  &     $-$1.67  &     0.06  &   18.28  &    0.02  &   16.96  &    0.01  &   16.44  &    0.01  &   16.24  &    0.01  &   16.13  &    0.01  &   30.7  \\
      53299$-$1961$-$148  &   323.4711914  &     $-$0.799086  &     $-$5.0  &      1.9  &      7696  &       74  &   3.31  &     0.20  &     $-$1.53  &     0.10  &   17.14  &    0.02  &   15.95  &    0.01  &   15.97  &    0.01  &   16.01  &    0.01  &   16.08  &    0.01  &   46.9  \\
      53299$-$1961$-$152  &   323.5119629  &     $-$0.501376  &      2.1  &      2.0  &      5176  &       38  &   2.38  &     0.26  &     $-$1.62  &     0.06  &   17.99  &    0.02  &   16.62  &    0.02  &   16.05  &    0.01  &   15.84  &    0.01  &   15.72  &    0.02  &   37.8  \\
\enddata

\end{deluxetable}
%%\end{document}

%\input{m2tab.tex}
\clearpage

%% Table 5
%%\documentclass[12pt,preprint]{aastex}
%%\pagestyle{empty}
\begin{deluxetable}{cccrccccccccccccccccccc}
\centering \tabletypesize{\scriptsize} \tablenum{7}
\tablecolumns{22} \tablewidth{0in} \rotate
\renewcommand{\tabcolsep}{2pt} \tablecaption{Properties of Selected Member Stars
of NGC~2420}
\tablehead{\colhead{spSpec name} & \colhead{RA} & \colhead{DEC} &
\colhead{RV} & \colhead{$\sigma_{\rm RV}$} & \colhead{T$_{\rm eff}$}
& \colhead{$\sigma_{\rm T_{\rm eff}}$} & \colhead{log $g$} &
 \colhead{$\sigma_{\rm log \it g}$} & \colhead {[Fe/H]} &
 \colhead{$\sigma_{\rm [Fe/H]}$} & \colhead{$u$} &
 \colhead{$\sigma_{u}$} & \colhead{$g$} & \colhead{$\sigma_{g}$} &
 \colhead{$r$} & \colhead{$\sigma_{r}$} & \colhead{$i$} &
 \colhead{$\sigma_{i}$} & \colhead{$z$} & \colhead{$\sigma_{z}$} &
 \colhead{$\rm <S/N>$} \\
\colhead{} & \colhead{(degree)} & \colhead{(degree)} & \colhead{(km
s$^{-1}$)} & \colhead{(km s$^{-1}$)} & \colhead{(K)} & \colhead{(K)}
& \colhead{(dex)} & \colhead{(dex)} & \colhead {(dex)} &
\colhead{(dex)} & \colhead{} & \colhead{} & \colhead{} & \colhead{}
& \colhead{} & \colhead{} & \colhead{} & \colhead{} & \colhead{} &
\colhead{} & \colhead{}}

\startdata

      53378$-$2078$-$101  &   114.7240067  &     21.321098  &     81.7  &      2.0  &    6191  &       36  &    4.13  &     0.17  &   $-$0.66  &     0.06  &      17.62  &    0.02  &   16.56  &    0.01  &   16.24  &    0.01  &   16.13  &    0.01  &   16.14  &    0.01  &   35.1  \\
      53378$-$2078$-$110  &   114.7332535  &     21.459303  &     87.1  &      1.8  &    6124  &       46  &    4.37  &     0.11  &   $-$0.57  &     0.18  &      17.68  &    0.02  &   16.56  &    0.01  &   16.23  &    0.01  &   16.12  &    0.01  &   16.11  &    0.01  &   38.2  \\
      53378$-$2078$-$111  &   114.7337723  &     21.533043  &     75.9  &      1.8  &    5887  &       36  &    4.44  &     0.09  &   $-$0.48  &     0.07  &      18.26  &    0.02  &   17.06  &    0.01  &   16.65  &    0.01  &   16.52  &    0.01  &   16.52  &    0.01  &   28.2  \\
      53378$-$2078$-$114  &   114.7632294  &     21.450647  &     74.0  &      1.7  &    6100  &       69  &    4.32  &     0.05  &   $-$0.56  &     0.12  &      17.82  &    0.02  &   16.70  &    0.01  &   16.37  &    0.01  &   16.24  &    0.01  &   16.28  &    0.01  &   36.1  \\
      53378$-$2078$-$116  &   114.7709503  &     21.494595  &     78.4  &      3.1  &    5117  &       77  &    4.54  &     0.07  &   $-$0.69  &     0.09  &      19.90  &    0.05  &   18.34  &    0.01  &   17.73  &    0.01  &   17.45  &    0.01  &   17.40  &    0.02  &   13.3  \\
      53378$-$2078$-$118  &   114.7212982  &     21.550411  &     74.6  &      2.6  &    5303  &       37  &    4.74  &     0.10  &   $-$0.66  &     0.08  &      19.56  &    0.04  &   18.05  &    0.01  &   17.47  &    0.01  &   17.27  &    0.01  &   17.21  &    0.02  &   16.6  \\
      53378$-$2078$-$120  &   114.7709274  &     21.527533  &     61.4  &      2.5  &    5529  &      107  &    4.16  &     0.09  &   $-$0.60  &     0.11  &      19.00  &    0.03  &   17.74  &    0.01  &   17.27  &    0.01  &   17.10  &    0.01  &   17.16  &    0.02  &   18.0  \\
\enddata
\end{deluxetable}
%%\end{document}

%\input{2420tab.tex}
\clearpage

%% Table 5
%%\documentclass[12pt,preprint]{aastex}
%%\pagestyle{empty}
\begin{deluxetable}{cccrccccccccccccccccccc}
\centering \tabletypesize{\scriptsize} \tablenum{8}
\tablecolumns{22} \tablewidth{0in} \rotate
\renewcommand{\tabcolsep}{2pt} \tablecaption{Properties of Selected Member Stars of
M~67} \tablehead{\colhead{spSpec name} & \colhead{RA} &
\colhead{DEC} & \colhead{RV} & \colhead{$\sigma_{\rm RV}$} &
\colhead{T$_{\rm eff}$} &
 \colhead{$\sigma_{\rm T_{\rm eff}}$} & \colhead{log $g$} &
 \colhead{$\sigma_{\rm log \it g}$} & \colhead {[Fe/H]} &
 \colhead{$\sigma_{\rm [Fe/H]}$} & \colhead{$u$} &
 \colhead{$\sigma_{u}$} & \colhead{$g$} & \colhead{$\sigma_{g}$} &
 \colhead{$r$} & \colhead{$\sigma_{r}$} & \colhead{$i$} &
 \colhead{$\sigma_{i}$} & \colhead{$z$} & \colhead{$\sigma_{z}$} &
 \colhead{$\rm <S/N>$} \\
\colhead{} & \colhead{(degree)} & \colhead{(degree)} & \colhead{(km
s$^{-1}$)} & \colhead{(km s$^{-1}$)} & \colhead{(K)} & \colhead{(K)}
& \colhead{(dex)} & \colhead{(dex)} & \colhead {(dex)} &
\colhead{(dex)} & \colhead{} & \colhead{} & \colhead{} & \colhead{}
& \colhead{} & \colhead{} & \colhead{} & \colhead{} & \colhead{} &
\colhead{} & \colhead{}}

\startdata
      54142$-$2667$-$361  &   132.6933289  &     11.871265  &     35.4  &      0.8  &     5350  &       96  &     3.99  &     0.29  &   $-$0.47  &     0.14  &    16.92  &    0.01  &   15.33  &    0.00  &   14.73  &    0.00  &   14.49  &    0.00  &   14.40  &    0.01  &   58.1  \\
      54142$-$2667$-$363  &   132.5890350  &     11.985747  &     32.4  &      0.8  &     5395  &      116  &     3.86  &     0.39  &   $-$0.20  &     0.28  &    16.78  &    0.02  &   15.26  &    0.01  &   14.75  &    0.02  &   14.57  &    0.01  &   14.56  &    0.01  &   57.3  \\
      54142$-$2667$-$364  &   132.6169128  &     11.913978  &     33.3  &      0.9  &     5855  &      167  &     4.45  &     0.17  &   $-$0.64  &     0.31  &    15.58  &    0.02  &   14.31  &    0.01  &   13.94  &    0.02  &   13.77  &    0.01  &   13.79  &    0.01  &   62.4  \\
      54142$-$2667$-$372  &   132.6952972  &     11.897903  &     33.9  &      0.8  &     5172  &      261  &     3.38  &     0.44  &   $-$0.59  &     0.07  &    17.33  &    0.01  &   15.67  &    0.00  &   15.07  &    0.00  &   14.86  &    0.01  &   14.80  &    0.01  &   54.0  \\
      54142$-$2667$-$378  &   132.6899872  &     11.926571  &     38.2  &      0.8  &     5079  &      124  &     4.83  &     0.15  &   $-$0.31  &     0.14  &    18.21  &    0.01  &   16.23  &    0.01  &   15.49  &    0.00  &   15.24  &    0.01  &   15.13  &    0.01  &   45.0  \\
      54142$-$2667$-$379  &   132.5899200  &     11.839747  &     35.4  &      0.9  &     5699  &      158  &     4.07  &     0.52  &   $-$0.33  &     0.14  &    15.96  &    0.01  &   14.66  &    0.02  &   14.23  &    0.01  &   14.13  &    0.02  &   14.09  &    0.02  &   61.4  \\
      54142$-$2667$-$402  &   132.7021637  &     12.014961  &     35.0  &      0.8  &     5551  &       43  &     3.41  &     0.65  &   $-$0.16  &     0.27  &    16.83  &    0.02  &   15.29  &    0.01  &   14.79  &    0.02  &   14.58  &    0.01  &   14.52  &    0.01  &   56.4  \\
\enddata
\end{deluxetable}
%%\end{document}

%\input{m67tab.tex}
\clearpage

\clearpage


\begin{thebibliography}{}

\bibitem[]{} Abazajian, K., et al. 2003, \aj, 126, 2081

\bibitem[]{} Abazajian, K., et al. 2004, \aj, 128, 502

\bibitem[]{} Abazajian, K., et al. 2005, \aj, 129, 1755

\bibitem[]{} Adelman-McCarthy, J. K., et al. 2007a, \apjs, in press

\bibitem[]{} Adelman-McCarthy, J. K., et al. 2007b, \apjs, accepted

\bibitem[]{} Allende Prieto, C., et al. 2007, \aj, submitted (Paper III)

\bibitem[]{} Allende Prieto, C., Beers, T. C., Wilhelm, R., et al.
             2006, \apj, 636, 804

\bibitem[]{} Bailer-Jones, C. A. L. 2000, \aap, 357, 197

\bibitem[]{} Beers, T. C., et al. 2006, BAAS, 38, 168.08

\bibitem[]{} Beers, T. C., Rossi, S., Norris, J. E., Ryan, S. G.,
             $\&$ Shefler, T. 1999, \apj, 506, 892

\bibitem[]{} Carollo, D., et al. 2007, Nature, submitted (astro-ph/0706.3005)

\bibitem[]{} Cenarro A.J., Cardiel N., Gorgas J., Peletier R.F., Vazdekis A., $\&$ Prada F. 2001a, \mnras, 326, 959

\bibitem[]{} Cenarro A. J., Gorgas J., Cardiel N., Pedraz S., Peletier R.F., $\&$ Vazdekis, A. 2001b, \mnras, 326, 981

\bibitem[]{} Chen, L., Hou, J.L., $\&$ Wang, J.J. 2003, \aj, 125, 1397,

\bibitem[]{} Cohen, J. G., Briley, M. M., $\&$ Stetson, P. B. 2005, \aj, 130, 1177

%\bibitem[]{} Cohen, J. G. $\&$ Mel$\acute{\rm e}$ndez, J. 2005, \aj, 129, 303

\bibitem[]{} Cudworth, K. M. 1976, \aj, 81, 519

\bibitem[]{} Cudworth, K. M. $\&$ Monet, D. G. 1979, \aj, 84, 774

\bibitem[]{} Cudworth, K. M. $\&$ Rauscher, B. 1987, \aj, 93, 856

\bibitem[]{} Dias, W. S., Alessi, B. S., Moitinho, A., $\&$ Lepine, J. R. D. 2002, \aap, 389, 871

\bibitem[]{} Fan, X., et al. 1996, \aj, 112, 628

\bibitem[]{} Friel, E. D.1989, \pasp, 101, 244

\bibitem[]{} Friel, E. D. $\&$ Janes, K. A. 1993, \aap, 267, 75

\bibitem[]{} Friel, E. D., Janes, K. A., Tavarez, M., Scott, J., et al. 2002, \aj, 124, 2693

\bibitem[]{} Fukugita, M., Ichikawa, T., Gunn, J.E., Doi, M., Shimasaku, K.,
             $\&$ Schneider, D.P. 1996, \aj, 111, 1748

\bibitem[]{} Girardi, L., Grebel, E. K., Odenkirchen, M., $\&$ Chiosi, C. 2004, \aap, 422, 205

\bibitem[]{} Gratton, R. 2000, in Stellar Clusters and Associations: Convection,
             Rotation, and Dynamos, ASP Conference Series
             (eds. R. Pallavicini, G. Micela, \& S. Sciortino), 198, p. 225

\bibitem[]{} Grillmair, C. J., Freeman, K. C., Irwin, M., $\&$ Quinn, P. J. 1995, \aj, 109, 2553

\bibitem[]{} Gunn, J. E., et al. 1998, \aj, 116, 3040

\bibitem[]{} Gunn, J. E., et al. 2006, \aj, 131, 2332

\bibitem[]{} Harris, W. E. 1996, \aj, 112, 1487

\bibitem[]{} Hogg, D.W., Finkbeiner, D.P., Schlegel, D.J., $\&$ Gunn, J.E. 2001,
              \aj, 122, 2129

\bibitem[]{} Ivezic, Z., et al. 2004, Astron. Nach., 325, 583

\bibitem[]{} Johnson, J. A. et al. 2007, in preparation

\bibitem[]{} Kraft, R. P. $\&$ Ivans, I. I. 2003, \pasp, 115, 143

\bibitem[]{} Lee, Y. S., et al. 2007a, \aj, submitted (Paper I)

\bibitem[]{} Lupton, R., et al. 2001, in ASP Conf. Ser. 238, Astronomical Data Analysis Software and Systems
             X, ed. F. R. Harnden, Jr., F. A. Primini, and H. E. Payne (San Francisco: Astr. Soc. Pac.), p. 269

\bibitem[]{} Morrison, H. L., Norris, J., Mateo, M., et al. 2003, \aj, 125, 2502

\bibitem[]{} Moultaka, J., Ilovaisky, S. A., Prugniel, P., $\&$ Soubiran, C. 2004, \pasp, 116, 693

\bibitem[]{} Otsuki, K., Honda, S., Aoki, W., Kajino, T., $\&$ Mathews, G. 2006, \apj, 641L, 117

\bibitem[]{} Pier, J.R., Munn, J.A., Hindsley, R.B., Hennessy, G.S.,
             Kent, S.M., Lupton, R.H., $\&$ Ivezic, Z. 2003, \aj, 125, 1559

\bibitem[]{} Prugniel, Ph. $\&$ Soubiran, C., 2001, \aap, 369,1048

\bibitem[]{} Randich, S., Sestito, P., Primas, F., Pallavicini, R., $\&$ Pasquini, L., 2006, \aap, 450, 557

\bibitem[]{} Rastorguev, A.S., Glushkova, E.V., Dambis, A.K., $\&$
             Zabolotskikh M.V. 1999, Astron. Letters, 25, 689

\bibitem[]{} Re Fiorentin, P., Bailer-Jones, C. A. L., Lee, Y. S. et al. 2007, \aap, 467, 1373

%\bibitem[]{} S\'anchez-Bl\'azquez, P., Peletier, R. F, Jim\'enez-Vicente, J., et al. 2006, \mnras, 371, 703

\bibitem[]{} Sanders, W. L. 1989, Rev., Mex. Astron. Astro. 17, 31

\bibitem[]{} Schlegel, D. J., Finkbeiner, D. P., $\&$ Davis, M., 1998, \apj, 500, 525

\bibitem[]{} Scott, J. E., Friel, E. D., $\&$ Janes, K. A. 1995, \aj, 109, 1706

\bibitem[]{} Smith, G. H $\&$ Briley, M. M. 2006, \pasp, 118, 740

\bibitem[]{} Smith, G. H. $\&$ Mateo, M. 1990, \apj, 353, 533

\bibitem[]{} Smith, J.A., et al 2002, \aj, 123, 2121

\bibitem[]{} Stetson, P. B. 1987, \pasp, 99, 191

\bibitem[]{} Stetson, P. B. 1990, \pasp, 102, 932

\bibitem[]{} Stetson, P. B. 1994, \pasp, 106, 250

\bibitem[]{} Stoughton, C., et al. 2002, \aj, 123, 485

\bibitem[]{} Twarog, B.A., Ashman, K.M., $\&$ Anthony-Twarog, B.J. 1997, \aj, 114, 2556

\bibitem[]{} Tucker, D., et al. 2006, AN, 327, 821

\bibitem[]{} Willemsen, P.G., Hilker, M., Kayser, A., $\&$
             Bailer-Jones, C. A. L. 2005, \aap, 436, 379

\bibitem[]{} Yong, D., Carney, B. W., $\&$ Teixera de Almeida, M. L. 2005, \aj, 130, 597

\bibitem[]{} York, D. G., et al. 2000, \aj, 120, 1579

\end{thebibliography}
\end{document}